\documentclass[preprint,12pt, 3p]{elsarticle}

\usepackage{amsmath}
\usepackage{amssymb}
\usepackage{mathrsfs}
\usepackage{hyperref}
\usepackage{bm}
\usepackage{accents}
\usepackage{xcolor}
\usepackage{graphicx}
\usepackage{wrapfig}
\usepackage[table]{xcolor}
\usepackage{color,soul}

\newdefinition{rmk}{Remark}

\journal{arXiv}

\begin{document}

\begin{frontmatter}

\title{Modeling cavitation and fibrillation in elastomers and adhesives. Part I: Cohesive instability}

\author[address1]{S. Mohammad Mousavi}
\author[address1]{Sarvesh Joshi}
\author[address4]{Franck Vernerey}
\author[address1,address5]{Nikolaos Bouklas\corref{corr}}
\ead{nb589@cornell.edu}
\address[address1]{Sibley School of Mechanical and Aerospace Engineering, \\ Cornell University, Ithaca, NY, USA}

\address[address4]{University of Colorado Boulder, Boulder, CO, USA}
\address[address5]{Pasteur Labs, Brooklyn, NY, USA }

\cortext[corr]{Corresponding authors}

\begin{abstract}
Cavitation in soft elastomers and adhesives is often viewed as an elastic instability often tied to the study of incompressible solids. It is the first step prior to fibrillation and ultimate failure in adhesives.  Building on the work of Lamont {\it{et al.}} (2025), elastomeric materials are treated as a crosslinked van der Waals fluid. The van der Waals contribution, capturing excluded volume and cohesive forces, is non-(poly)convex, readily providing an intrinsic analytical criterion for cavity nucleation.  This work introduces a gradient-enhanced continuum framework that examines the emergence of cavity formation from the perspective of a cohesive instability and corresponding phase transition without requiring a pre-existing defect. The corresponding thermodynamically consistent derivation includes the introduction of a relevant material length scale as well as viscous dissipation associated with polymer chain disentanglement during the cohesive instability. This work does not study the impending damage that the material undergoes during the cohesive instability and transition from a dense to a rare phase. Interestingly, it is shown that for both strain stiffening and strain softening models (in terms of their shear response), an instability reminiscent of what is expected in the case of cavitation is recapitulated. Simulations reproduce key experimental trends, includingthe aspect ratio–driven transition from few large to many small cavities depending on the thickness of an adhesive layer. The framework offers a robust, physically grounded basis for the cohesive instability that drives cavity nucleation, enabling future integration with damage, fracture, and dissipation models to capture the complete cavitation, fibrillation, and failure process. 
\end{abstract}


\end{frontmatter}

\section{Introduction}\label{Section:Intro}
Cavitation, the emergence of optically visible cavities under applied stress, is an early signature for triggering the failure cascade of rubber-like materials \citep{lakrout1999direct, creton2016fracture, barney2020cavitation}.  As the voids expand, the material ligaments between them are drawn into thin, load-bearing strands or fibrils often encountered in adhesive \citep{lakrout1999direct, zosel1989adhesive, varchanis2021adhesion}. This fibrillation stage is critical, as the large-scale plastic drawing of these fibrils dissipates a significant amount of energy, which is the primary source of the material's adhesion energy \cite{zosel1998effect, yang2022rate}. The phenomenon of cavitation, which is at the core of this sequence of events, has been a subject of extensive research since the seminal experiments of Gent and Lindley (1959) \cite{gent1959internal}.   Theoretical work on the topic has focused on both the unstable growth of pre-existing defects as well as the formation and growth of cavities. Green and Zerna (1954) \cite{green1992theoretical} were the first to present a solution for the problem of a hollow spherical shell, similar to the case of a pre-existing defect, and Williams and Shapery \cite{williams1965spherical} treated the problem of the cavity growth from an energetic standpoint, connecting it to concepts from fracture mechanics. The pioneering work of Ball (1982) \cite{ball1982discontinuous} focused on the emergence of a cavity in a homogeneous and isotropic (incompressible and compressible) elastic sphere upon reaching a critical remote load, through the investigation of a class of bifurcation problems for nonlinear elasticity assuming spherical symmetry and leading to discontinuous deformations.  The interested reader is directed to the review by Horgan and Polignone (1995) \cite{horgan1995cavitation} and citations therein, which focuses on theoretical results established using the bifurcation approach of Ball, but also other works that focus on the case of pre-existing defects, extensions to the compressible case, anisotropic and non-spherically symmetric cases, among others. The effect of surface stresses on the cavity boundary was taken into account in Gent and Tompkins (1969) \cite{gent1969surface} in rubber-like solids, but also more recently in Dollhofer {\it{et al.}} (2004) \cite{dollhofer2004surface}. The latter tries to bridge the gap between experimental observations and the original predictions of Ball, focusing on questions related to size effects on cavitation. This series of works, following the seminar paper of Ball do not explicitly connect to microscopic and statistical characteristics of polymeric networks and, as such leave several unanswered questions regarding the connection of molecular and network characteristics to eventual response.

Other important reviews on the topic include those from Gent (1990) \cite{gent1990cavitation} and  Fond (2001) \cite{fond2001cavitation}. The perspective from Barney {\it{et al.}} (2019) also highlights the field of cavitation rheology, and the relevance of cavitation phenomena in a wide range of fields such as biological materials. Needle-induced cavitation is also discussed in the latter, where one needs to note that the boundary conditions and geometry of the associated problem are significantly different from the ones discussed above. More recent works extend to complex loading scenarios, including generalized loading conditions \citep{lopez2011cavitation}, examine defect growth at material interfaces, a mechanism termed interfacial cavitation \citep{henzel2022interfacial}, but also examine the competition of elastic cavitation and fracture mechanisms  \cite{hutchens2016elastic,raayai2019intimate}. There are several works that have focused on interpreting the multiscale nature of cavitation as damage without explicitly transitioning to fracture, including \cite{dorfmann2002shear}, and also from a computational multiscale perspective in \cite{dal2013multiscale}, where the lower scale included pre-existing voids.  The connection of cavitation and fracture is intimate, as was noted in Irwin \cite{Irwin1958}, discussing the connection of cavitation events in fluids to fracture in solids as fundamental mechanisms of surface creation. More recently, this question has been revisited in the context of cavitation of soft materials \cite{lefevre2015cavitation, poulain2017damage}, with the emergence of a body of work extending classical phase field models for brittle fracture \cite{borden2012phase,miehe2010rate}. These models were unable to capture crack nucleation (beyond a uniaxial stress state). Works including  Kumar \textit{et a}l. (2020) \cite{kumar2020revisiting} and Kamarei \textit{et al.} (2025) \citep{kamarei2025nucleation} have incorporated a strength surface to capture crack nucleation; in this approach, and upon their emergence, cavities are numerically captured as cracks with a distinct front.  

Towards this direction, two primary schools of thought have emerged. The first seeks to preserve a single-functional variational framework through increasingly sophisticated energy decompositions. Notable contributions include those by De Lorenzis and Maurini (2022) \cite{de2022nucleation} and, more recently, Vicentini \textit{et al.} (2025) \cite{vicentini2025variational}, whose cohesive fracture model enables flexible tuning of the strength surface. The second, alternative approach introduces material strength directly into the phase-field evolution equation via an external driving force. Originally proposed by Kumar \textit{et al.} (2018) \cite{kumar2018fracture}, this formulation was initially critiqued for its non-variational character. However, Larsen \textit{et al.} (2024) \cite{larsen2024variational} later resolved this by reformulating the theory as a variational principle based on the alternating minimization of two distinct functionals: one governing deformation and the other fracture. This two-functional variational framework was subsequently generalized by Chockalingam (2025) \cite{chockalingam2025construction}, who developed a systematic procedure to construct the driving force for arbitrary strength surfaces, such as Mohr–Coulomb. This comprehensive phase-field formulation has successfully reproduced the classic poker-chip experiments \cite{kumar2021poker} and has been extended to dynamic fracture, where material strength remains crucial for accurately predicting impact-driven failure \cite{liu2024effects}. Moreover, it naturally captures the inherent tension–compression asymmetry of fracture without requiring artificial energy splits \cite{liu2025emergence}. Altogether, these advances establish a robust and physically grounded framework capable of simulating both crack nucleation and propagation in soft materials. One clear limitation is the lack of information from lower scales on the construction of the aforementioned strength surfaces, which are mainly motivated from a phenomenological perspective.


While these advanced phase-field frameworks provide a robust methodology for incorporating material strength, the physical origins of this strength surface remain an open question,  as well as the potential transitions from a spherically symmetric growth of a cavity accompanying diffuse damage along the cavity boundary to that of a fracture-like growth of a pre-existing cavity or defect.  Using a novel phase-separation technique to controllably grow cavities from the nano- to micro-scale, Kim \textit{et al.} (2020) \cite{kim2020extreme} observed that initial cavity expansion occurs at a constant pressure and involves irreversible bond breakage that is distributed around the cavity, drawing strong analogies to ductile void growth in metals rather than classical brittle fracture. This suggests that while macroscopic failure may appear brittle, the initial stages of damage can involve significant distributed inelasticity. Molecular dynamics (MD) simulations suggest a fundamental challenge to classical theories: above the glass transition temperature, elastomers typically exhibit only nano-sized defects, not the micron-sized flaws assumed in many continuum models \cite{zee2015cavitation}. This implies that cavitation may initiate at the molecular level, likely when local stresses overcome the cohesive strength of intermolecular forces, such as van der Waals interactions. Indeed, coarse-grained MD simulations confirm that the pressure required for initial cavitation significantly exceeds predictions based solely on continuum elastic instability \cite{ye2020molecular}. Recent theoretical efforts have begun to address this gap by deriving failure criteria based on fundamental physics. For instance, Lamont \textit{et al.} (2025) \cite{lamont2025cohesive} proposed a model depicting elastomers as a crosslinked van der Waals fluid. Noting van der Waals' work on liquid-vapor interfaces \cite{van1979thermodynamic}, the approach of Lamont reveals the emergence of a cohesive instability where failure does not initiate from a pre-existing flaw. Such a framework offers a pathway to connect molecular physics directly to the macroscopic strength criteria needed for continuum fracture models, bridging a critical gap between scales. Notably, the model by Lamont, which is at the core of this study, considers the solid nature of elastomers coming from a cross-linked network of chains that provides a resistance to shear, but also the fluidic nature of this class of materials, where van der Waals interactions between monomers (or Kuhn segments) provide resistance to bulk deformations. Conceptually, this approach views a phase-transition at the core of cavity nucleation without assuming pre-existing defects and deviating from the approach of Ball, but is conceptually adjacent to early theoretical works of Podio-Giudugli and co-workers \cite{podio1987hyperelastic, lancia1996gleanings} which, though, lack the explicit connection to statistical mechanics of polymer chains.

Taking the model of Lamont \textit{et al.} (2025) \cite{lamont2025cohesive} as a starting point, this study seeks to obtain numerical solutions for the cohesive instability that precedes cavity formation and growth, as well as fibrillation and ultimate failure. This work does not yet introduce a damage theory, and as such, there is no treatment for polymer chains that might rupture during the cohesive instability; this will be the topic of follow-up work in Part II. In this context, gradient-enhanced models could serve as a suitable nonlocal continuum damage tool that allows for the integration of information from statistical mechanics while accounting for network-level heterogeneity. This is achieved through an integrated framework that combines network-level modeling, statistical mechanics, and continuum-scale simulations \cite{li2020variational, mousavi2026capturing, joshi2026instabilities}. Viscoelastic contributions, which are especially important for the case of adhesives, will be the focus of Part III.  The remainder of this paper is organized as follows: Section \ref{Section:theory} presents the theoretical preliminaries and the developed gradient-enhanced model for capturing cohesive instabilities in elastomers. Section \ref{Section:NumericalImplementation} details the thermodynamically consistent formulation of a numerical framework and corresponding implementation using the open-source finite element library FEniCS \cite{alnaes2015fenics}. In Section \ref{Section:results}, the model is employed to simulate cavitation across two boundary value problems, encompassing five representative case studies. The complete implementation is available on GitHub\footnote{\url{https://github.com/MMousavi98/cavitation}}.
\section{A Gradient-Enhanced Theory}\label{Section:theory}

\subsection{Kinematics}\label{SubSection:Kinematics}

The reference configuration of the body, ${\Omega}_0 \subset \mathbb{R}^3$ is considered as an open, bounded, and connected domain with a sufficiently smooth boundary, denoted by $\partial \Omega_0$. As the body undergoes deformation, each material point moves from its initial position $\bm{X}$ in the reference configuration $\Omega_0$ to a new position $\bm{x}$ in the deformed configuration $\Omega$. The displacement vector, defined as $\bm{u} = \bm{x} - \bm{X}$, characterizes this change in position from the reference to the current state. A complete description of the deformation requires analyzing how infinitesimal line elements are transformed, which is expressed through the relation:
\begin{equation}\label{bulk-element}
d\bm{x} = \mathbf{F}(\bm{X},t)\, d\bm{X},
\end{equation}
where $\mathbf{F}$ is the deformation gradient, given by
\begin{equation}\label{deformation-gradient}
\mathbf{F}(\bm{X},t) = \frac{\partial \bm{x}}{\partial \bm{X}}.
\end{equation}
The Jacobian determinant $J(\bm{X}, t) = dv/dV = \det \mathbf{F}(\bm{X}, t)$ measures the local change in volume between the reference and deformed configurations and must remain positive to ensure physically admissible deformations. In addition, the right Cauchy-Green deformation tensor $\mathbf{C}$ is introduced as
\begin{equation}
\mathbf{C} = \mathbf{F}^T \mathbf{F},
\end{equation}
whose principal invariants are defined as follows \cite{holzapfel2002nonlinear}:
\begin{equation} \label{invariants}
\begin{split}
    I_1 = \mathrm{tr}(\mathbf{C}) \\
    I_2 = \frac{1}{2}\left[(\mathrm{tr}(\mathbf{C}))^2 - \mathrm{tr}(\mathbf{C}^2)\right]\\
    I_3 = \det(\mathbf{C}),
\end{split}
\end{equation}
where $J=det\mathbf{F}=\sqrt{I_3}$.

\subsection{Hyperelasticity and the condition of incompressiblity}\label{hyperelasticity}
In this work, we consider a homogeneous, isotropic, and incompressible hyperelastic material subjected to quasi-static loading. To describe its mechanical response, we introduce the Helmholtz free energy function $\psi$, defined per unit reference volume, which forms the basis for deriving the constitutive relations. Different hyperelastic models capture distinct aspects of material behavior; accordingly, we examine two forms of the free energy in this study: the classical Neo-Hookean model, representing a baseline elastic network, and the monodisperse network model, which provides a more detailed account of chain-level mechanics.  
For simplicity, we first examine the Neo-Hookean free energy density, which is widely used to model soft materials such as elastomers and biological tissues, while acknowledging its limitations in accurately representing elastomers beyond moderate chain stretches.
\begin{equation}\label{neo-hookean}
{\psi}(\mathbf{F}) =\frac{\mu}{2}(I_1-3),
\end{equation}
where $\mu$ is the shear modulus. 

In various experimental studies, the near-incompressibility of elastomers, polymers, and biological tissues has been confirmed \cite{holzapfel2002nonlinear}, and enforcing this constraint (along with alleviating the numerical issues that arise when enforcing it) has been the focus of an extensive body of work \cite{wriggers2021taylor}. The constraint can be exactly enforced through the introduction of a Lagrange multiplier $p$. This leads to the following form for the free energy density
\begin{equation}\label{psi-lagrange}
\psi(\mathbf{F},p) = \frac{\mu}{2}(I_1-3 -2 \mathrm{ln}J) -p\left(J -1\right)=\psi_0(\mathbf{F})-p\left(J -1\right).
\end{equation}
Turning to a penalty formulation adds an energetic contribution for volumetric deformations, leading to
\begin{equation}\label{psi-penalty}
\psi(\mathbf{F}) = \psi_0(\mathbf{F}) +\frac{\kappa}{2}\left(J -1\right)^2, \quad \mathrm{or} \quad \psi(\mathbf{F}) = \psi_0(\mathbf{F}) +\frac{\kappa}{2}\left(\mathrm{ln}J\right)^2
\end{equation}
along with other common forms. This allows us to approach the condition of near-incompressibility for high values of the ratio $\kappa/\mu$, but leads to several issues when used in numerical implementations \cite{wriggers2008nonlinear, auricchio2013approximation}.
Alternatively, a perturbed Lagrangian formulation can be utilized \cite{ ang2022stabilized, brink1996some, li2020variational}.  Through this approach, we can express the energy density function as follows
\begin{equation}\label{psi-pertL}
\psi(\mathbf{F},p) = \psi_0(\mathbf{F}) -p\left(J -1\right) -\frac{p^2}{2\kappa}.
\end{equation}
where $\kappa$ controls the degree of compressibility. 

\subsection{Constitutive models based on polymer chain statistics }

 Similar to how the Neo-Hookean model can be derived from principles of statistical mechanics for polymer chains restricted to moderate chain deformations \cite{destrade2017methodical}, models that can provide a more realistic description of the material response at the limit of chain rupture have been developed \cite{Arruda-Boyce1993}. Such models have previously been enhanced to simulate network fracture and damage at the continuum level \cite{li2020variational, talamini2018progressive,mousavi2025chain, mousavi2026capturing}. Following these works, we consider a non-Gaussian formulation \cite{james1943theory, treloar1975physics} that accounts for the finite extensibility of polymer chains as the end-to-end distance $r$ approaches the contour length $Nb$, where $N$ is the number of Kuhn segments and $b$ is the segment length. The Helmholtz free energy of an individual chain composed of $N$ freely jointed segments of length $b$ is given by \citep{james1943theory, treloar1975physics, flory1953principles}:

\begin{equation}
\label{free_energy_single_chain}
\psi_{entropic}(r) = k_bT N\left(\frac{r}{Nb}\beta+\ln\frac{\beta}{\sinh\beta}\right), \quad \beta = \mathscr{L}^{-1}\left(\frac{r}{Nb}\right),
\end{equation}
where $T$ is temperature and $k_b$ is the Boltzmann constant.

Polymer network models grounded in the statistical mechanics of long-chain molecules \cite{treloar1979non, arruda1993three, wu1993improved, boyce2000constitutive} have successfully captured experimental responses under various large-strain deformation modes, including uniaxial tension, equibiaxial extension, and pure shear. These models employ a non-Gaussian chain description, in which the chain stretch approaches its contour length, causing the free energy \eqref{free_energy_single_chain} to diverge. As a result, their validity is confined to the regime $r < Nb$, preventing them from representing chain scission. To overcome this limitation, \cite{mao2017rupture} extended the classical framework by incorporating ideas from \cite{smith1996overstretching}, originally developed for modeling DNA overstretching. This approach relaxes the rigidity assumption of Kuhn segments, accounting for both segmental deformation and alignment under tension. In doing so, the free energy of a single chain is augmented with an enthalpic contribution associated with stretching the Kuhn segments, thereby enabling a reasonable approximation of the mechanical behavior up to the point of chain scission. The chain Helmholtz free energy can thus be expressed as
\begin{align}
\label{free_energy_single_chain_extend0}
\psi_{chain}\left(r,\lambda_\mathrm{b}\right) = U(\lambda_\mathrm{b}) + k_bT N\left(\frac{r}{N \lambda_\mathrm{b}b}\beta+\ln\frac{\beta}{\sinh\beta}\right), \quad \beta 
= \mathscr{L}^{-1}\left(\frac{r}{N \lambda_\mathrm{b}b}\right), 
\end{align}
where $\lambda_\mathrm{b}$ represents the Kuhn segment stretch ratio, and $U(\lambda_\mathrm{b})$ is the internal energy stored within the stretched Kuhn segments. For an individual polymer chain, the chain stretch $\lambda_\mathrm{ch}$ is defined as the ratio of its deformed length $r$ to the equilibrium length $r_0$, i.e., $\lambda_\mathrm{ch} = r / r_0$. Accordingly, based on \eqref{free_energy_single_chain_extend0}, the chain’s free energy can be written as a function of $\lambda_\mathrm{ch}$ and $\lambda_\mathrm{b}$:
\begin{equation}
\label{free_energy_single_chain_extend}
\psi_{chain}\left(\lambda_\mathrm{ch},\lambda_\mathrm{b}\right) = U(\lambda_\mathrm{b}) + k_bTN\left(\frac{\lambda_\mathrm{ch}\lambda_\mathrm{b}^{-1}}{\sqrt{N}}\beta+\ln\frac{\beta}{\sinh\beta}\right), \quad \beta = \mathscr{L}^{-1}\left(\frac{\lambda_\mathrm{ch}\lambda_\mathrm{b}^{-1}}{\sqrt{N}}\right).
\end{equation}

By comparing the extensible free energy functional \eqref{free_energy_single_chain_extend} with the classical form \eqref{free_energy_single_chain}, it becomes clear that the modified stretch term $\lambda_\mathrm{ch}\lambda_\mathrm{b}^{-1}$ represents the portion of the chain stretch attributable exclusively to the reorientation and alignment of Kuhn segments, independent of bond stretching effects \cite{talamini2018progressive, mao2017rupture}. Following this, we employ a simple quadratic expression for the chain’s internal energy, as introduced in \cite{talamini2018progressive, mao2018theory}:
\begin{equation}
\label{single_chain_internal_energy}
U(\lambda_\mathrm{b}) = \frac{1}{2} N E_\mathrm{b}\left( \lambda_\mathrm{b} - 1\right)^2,
\end{equation}
where $E_\mathrm{b}$ denotes the stiffness of the chemical bonds in the chain backbone.

Additionally, using the 8-chain model proposed by Arruda and Boyce \cite{Arruda-Boyce1993}, the average chain stretch at a given material point in the elastomer can be represented as
\begin{equation}\label{Lambda_chain}
\lambda_\mathrm{ch} = \sqrt{\frac{I_1}{3}},
\end{equation}
assuming that all chains contain the same number of segments. Finally, by extending the single-chain free energy \eqref{free_energy_single_chain_extend} to the network level and introducing the chain density $n$ (number of chains per unit reference volume), the Helmholtz free energy density of the elastomer network is obtained as:
\begin{equation}
\label{psi_iso}
\psi_{net}\left(\mathbf{F}\right) = \frac{1}{2} N E \left(\lambda_\mathrm{b} - 1\right)^2 + N\mu\left(\frac{\lambda_\mathrm{ch}\lambda_\mathrm{b}^{-1}}{\sqrt{N}} \beta + \ln\frac{\beta}{\sinh\beta}\right), \quad \beta = \mathscr{L}^{-1}\left(\frac{\lambda_\mathrm{ch}\lambda_\mathrm{b}^{-1}}{\sqrt{N}}\right),
\end{equation}
where $\mu = nk_bT$ is the shear modulus of the network, and $E = nE_\mathrm{b}$ represents the stiffness associated with bond stretching. 
To account for near-incompressibility, one could at this point utilize the penalty of the perturbed Lagrangian formulation, where the latter would lead to: 
\begin{equation}\label{psi-chain-pertL}
\psi(\mathbf{F}) =\psi_{net}\left(\mathbf{F}\right) -p\left(J -1\right) -\frac{p^2}{2\kappa}.
\end{equation}
Even though similar expressions are widely used, there is an inherent discrepancy in such formulations, as employing polymer chain statistics does not naturally enforce near-incompressibility, and the ad-hoc addition from Eqs. \eqref{psi-lagrange}, \eqref{psi-penalty}, and \eqref{psi-pertL} are utilized to enable constrained optimization.

\subsection{A crosslinked van der Waals fluid model}
To provide a more physically-grounded description, while considering the different microscopic origins of shear and volumetric resistance in elastomers, we adopt the modeling approach introduced by Lamont \textit{et al.} (2025) \cite{lamont2025cohesive} and model the elastomer as a crosslinked van der Waals fluid. This enables taking into account crosslinking and chain statistics, but also considers the non-covalent interactions between individual monomers. Prior to crosslinking, a polymer melt is nearly incompressible; this attribute is due to the non-covalent interactions between individual monomers, and as such, a reasonable description is that of a van der Waals fluid \cite{lamont2025cohesive, tabor1994bulk}.  The model of Lamont \textit{et al.} (2025) \cite{lamont2025cohesive} (in general) separates the total free energy into two distinct contributions: a network elasticity contribution (that could be either Neo-Hookean from Eq. \eqref{neo-hookean} or extended to follow Langevin statistics and enthalpic contributions from Eq. \eqref{psi_iso}) and a van der Waals fluid contribution. This approach enables capturing the cohesive interactions among polymer segments in the densely packed bulk of the elastomer that are enthalpic and entropic in origin. Additionally, it departs from traditional assumptions of the enforcement of perfect and near-incompressibility.

The free energy density corresponding to the van der Waals fluid description is defined as a function of the volume ratio $J$ as:
\begin{equation}\label{psi-vol}
\psi_{VW}(J) = -\frac{nk_{b}T}{\chi}\left[\ln(J-f_{0})+\frac{\epsilon_{a}}{k_{b}T}\frac{f_{0}}{J}\right],
\end{equation}
where $\epsilon_a$ is the average attractive potential energy between ``fluid" particles, here thought of as monomers or Kuhn segments. The term $f_0$ represents the initial volume fraction occupied by particles in the reference state, where $f_0 = V_p / V_0 \leq 1$, and $V_p$ is the total volume occupied by particles ($n\nu = V_p$). The parameter $\chi$ is defined as the ratio of the number of polymer chains to the number of particles, calculated as $V_0 c_0 / n$. A reasonable approximation for $\chi$ is $1/N$, which assumes each Kuhn segment of a polymer chain in the network acts as a volume-occupying particle in the van der Waals fluid.

Eq. \eqref{psi-vol} has two components that account for intermolecular interactions. The first part, characterized by the logarithmic term $\text{ln}(\Bar{J}-f_0)$, represents the effect of excluded volume. This term acknowledges that the constituent particles (e.g., Kuhn segments) have a finite size, which reduces the available volume for movement and contributes entropically to the free energy. The second part of the equation, which includes the attraction energy parameter $\epsilon_a$, accounts for the existence of attractive molecular forces between monomers. This term is an enthalpic contribution representing the cohesive energy that maintains the integrity of the polymer segments against expansion. Together, the interplay between the entropic repulsion from excluded volume and the enthalpic cohesion from attractive forces dictates the material's bulk modulus and its stability against volumetric deformation. 

Finally, we obtain an expression for the free energy density utilizing a compressible Neo-Hookean model from Eq.\eqref{psi-lagrange} and the van der Waals fluid contribution from Eq.\eqref{psi-vol} as:
\begin{equation}\label{psi-iso-VW1}
\psi_{NH-VW}(\mathbf{F}) = \psi_\mathrm{0}(\mathbf{F})+\psi_{VW}(J)= \frac{\mu}{2}(I_1-3-2\ln J) -\frac{nk_{b}T}{\chi}\left[\ln(J-f_{0})+\frac{\epsilon_{a}}{k_{b}T}\frac{f_{0}}{J}\right].
\end{equation}
\begin{figure}[h!]
    \centering
    \includegraphics[width=0.5\linewidth]{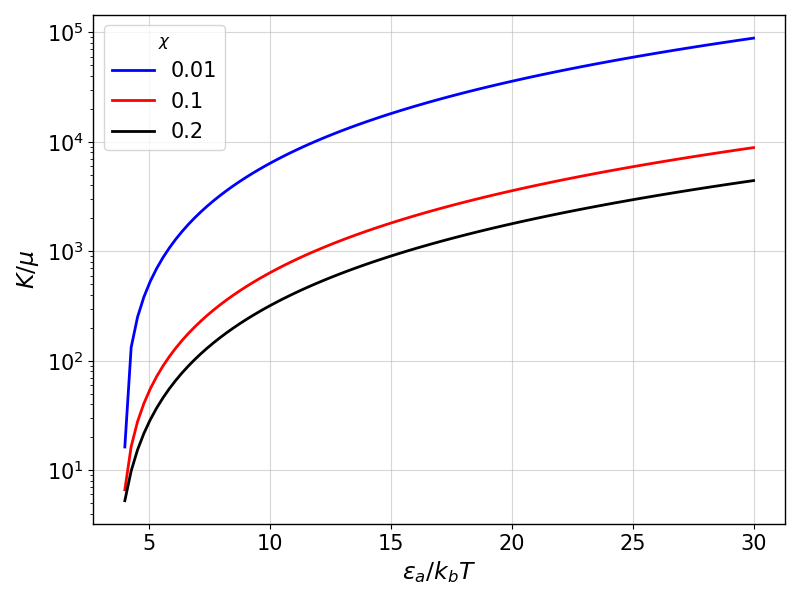}
    \caption{Normalized bulk modulus considering $\mu=1$.}
    \label{fig:bulk-modulus}
\end{figure}
The corresponding bulk modulus obtained from this model is plotted in Fig. \eqref{fig:bulk-modulus}, and the details of the calculation can be found in \ref{appendix1}. Alternatively, by utilizing the more advanced model derived based on chain statistics allowing for chain extensibility from Eq. \eqref{psi_iso} in conjunction with the van der Waals contribution, we obtain:
\begin{align}\label{psi-iso-VW2}
\psi_{net-VW}(\mathbf{F}) &= \psi_{net}(\mathbf{F})+\psi_{VW}(J) \\
&= \frac{1}{2} N E \left(\lambda_\mathrm{b} - 1\right)^2 + N\mu\left(\frac{\lambda_\mathrm{ch}\lambda_\mathrm{b}^{-1}}{\sqrt{N}} \beta + \ln\frac{\beta}{\sinh\beta}\right) -\frac{nk_{b}T}{\chi}\left[\ln(J-f_{0})+\frac{\epsilon_{a}}{k_{b}T}\frac{f_{0}}{J}\right]. \nonumber
\end{align}
For both models in Eqs.\eqref{psi-iso-VW1} and \eqref{psi-iso-VW2}, the solid and fluidic nature of an elastomeric material is captured, with near-incompressibility arising naturally as can be seen in Fig. \eqref{fig:bulk-modulus} following the effective bulk-to-shear-modulus ratio (depicted through linearization at the undeformed configuration for a range of material parameters).  The models in Eqs.\eqref{psi-iso-VW1} and \eqref{psi-iso-VW2}, are extensively utilized in the remainder of this study.

Additionally, it is important to note that, as seen in Lamont \textit{et al.} (2025) \cite{lamont2025cohesive}, these non-polyconvex energies can allow for instabilities during expanding volumetric deformations. These cohesive instabilities (as we will refer to them onwards) are reminiscent of cavity formation in fluids \cite{wang2017cavitation}, which are captured in a continuum setting with van der Waals fluid models and are accompanied with phase transition from a dense to a dilute phase (liquid-vapor phase transition) \cite{liu2015liquid}, but also reminiscent of cavitation in solids \cite{ball1982discontinuous,dollhofer2004surface}. We note that in this work, a phase transition from a dense to a dilute phase would not be accommodated with damage, as we have not included such a mechanism for chain scission (this will be the focus of Part II of this work). Commonly, cavitation in solids is studied in the presence of a defect (void), but similar to fluids in this model, thermal fluctuations are expected to trigger a cohesive instability event.

\subsection{A gradient-enhanced model}
The local models presented up to this point fail to capture phenomena that are associated with emergent length scales due to the material microstructure. Such effects in elastomers might arise due to network architecture/connectivity, network imperfections and chain polydispersity \cite{creton2016fracture, LIN2021101399}, and have a significant impact in fracture and damage processes \cite{Irwin1958, mousavi2026capturing, chen2017flaw}; the ratio of the critical energy release rate $G_c$ and rupture energy $W^*$ (in uniaxial tension), is often refereed to as the fractocohesive length scale $\ell_{fc}=G_c/W^*$. Additionally, surface and interfacial tension can often dominate the response in soft materials, especially when the elastocapillary length scale (the ratio of surface/interface tension to shear modulus) $\ell_{ec}=\gamma/\mu$, is comparable or larger than the characteristic feature size. It is noted that both length scales can be at the $\sim$mm level \cite{long2021fracture, bico2018elastocapillarity}.

To introduce a characteristic length scale in our formulation, we resort to a (partial) gradient elasticity approach, where instead of penalizing the gradient of the full deformation gradient, we penalize the material gradient of volume ratio $J$. Thus, we formulate a  gradient penalization as:
\begin{equation} \label{psi_grd}
    \psi_{grd}(\nabla J) = \frac{\ell^2}{2} \nabla J \cdot \nabla J,
\end{equation}
where $\ell$ is a material length scale that controls interface formation, penalizing the gradient of the volume ratio. The models from Eqs.\eqref{psi-iso-VW1}-\eqref{psi-iso-VW2} can be reformulated to include gradient contributions as:
\begin{align}\label{psi-iso-VW3}
    &\psi_{NH-VW}(\mathbf{F},\nabla J) = \psi_\mathrm{0}(\mathbf{F})+\psi_{VW}(J)+\psi_{grd}(\nabla J)  \\
    &\psi_{net-VW}(\mathbf{F},\nabla J) = \psi_{net}(\mathbf{F})+\psi_{VW}(J)+\psi_{grd}(\nabla J). \label{psi-iso-VW4}
\end{align}

\section{Numerical Implementation}
As the model presented in Section \ref{Section:theory} would require $C^1$ continuity, we further develop a framework that requires $C^0$ continuity and is thermodynamically consistent.

\subsection{Thermodynamic consistency}
To enable a straightforward numerical implementation, we introduce an additional scalar field $\Bar{J}$, serving as a nonlocal counterpart to the local volume ratio J. As such, the models developed in Eqs.\eqref{psi-iso-VW3}-\eqref{psi-iso-VW4} can be recast as:
\begin{align}
    &\psi_{NH-VW}(\mathbf{F},\Bar{J},\nabla \Bar{J}) = \psi_\mathrm{0}(\mathbf{F})+\psi_{VW}(\Bar{J}) + \psi_{grd}(\nabla \Bar{J}) +\psi_{coup}(J, \Bar{J}) \label{psi-iso-NHVW5} \\
    &\psi_{net-VW}(\mathbf{F},\Bar{J},\nabla \Bar{J}) = \psi_{net}(\mathbf{F})+\psi_{VW}(\Bar{J}) + \psi_{grd}(\nabla \Bar{J}) +\psi_{coup}(J, \Bar{J}), \label{psi-iso-StVW5}
\end{align}
noting that the van der Waals and gradient parts above are now in terms of the nonlocal volume ratio, contrary to how they were originally presented in the previous section. The coupling term $\psi_{coup}$ introduced above, penalizes deviations between the local and nonlocal volume ratios as:
\begin{equation} \label{psi_coup}
    \psi_{coup}(J, \Bar{J}) = c \left[J - \Bar{J}\right]^2,
\end{equation}
where $c$ is a coupling modulus (this is a penalty-based approach to approximate $J=\Bar{J}$, and as such $c$ needs to take sufficiently large values to approximately enforce the constraint, but at the same time not large enough to impede convergence; alternately one could introduce a Lagrange multiplier, or a perturbed Lagrangian scheme). 

Further, the principle of virtual power is employed to derive the governing equations. The internal mechanical power $P_{int}$ for a nearly incompressible polymer network in the reference configuration $\Omega_0$ is given by:
\begin{equation}\label{rateIntPower}
    P_{int} = \int_{\Omega_0}\left[\mathbf{P}\colon\nabla\dot{\bm{u}} + f_{\Bar{J}}\dot{\Bar{J}} + \boldsymbol{\xi}_{\Bar{J}}\cdot\nabla\dot{\Bar{J}}\right]dV,
\end{equation}
Here, $\mathbf{P}$ denotes the first Piola–Kirchhoff stress tensor, while $f_{p}$, $f_{\Bar{J}}$, and $\boldsymbol{\xi}_{\Bar{J}}$ are the power conjugates associated with the internal state variable $\Bar{J}$ and its gradient. Neglecting body forces, the external mechanical power $P_{ext}$ is expressed as:
\begin{equation}
    P_{ext} = \dot{W}_{ext} = \int_{\partial_N\Omega_0}{\bm{T}}\cdot\dot{\bm{u}} \,dA.
\end{equation}
where $\bm{T}$ denotes the mechanical surface traction. Enforcing the principle of virtual power (i.e., the first law of thermodynamics in variational form, $\delta P_{int} = \delta P_{ext}$) and applying the divergence theorem yields the following governing equations along with the associated boundary conditions:
\begin{itemize}
    \item Mechanical equilibrium equation and boundary conditions: \begin{equation} \label{mechanical_equilibrium_governing}
    \begin{split}
        \nabla\cdot\mathbf{P}= \mathbf{0} \quad \text{in} \quad \Omega_0,\\
        \bm{u}= \Bar{\bm{u}} \quad \text{on} \quad \partial_{D}{\Omega}_0,\\
        \mathbf{P}\cdot\bm{n}_0 = \bm{T} \quad \text{on} \quad \partial_N\Omega_0.
        \end{split}
    \end{equation}

    \item Micro-force equilibrium equation and boundary conditions: \begin{equation} \label{chain_governing}
    \begin{split}
        \nabla\cdot\boldsymbol{\xi}_{\Bar{J}}-f_{\Bar{J}} = 0 \quad \text{in} \quad \Omega_0,\\
        \Bar{J}= \Bar{\Bar{J}} \quad \text{on} \quad \partial_{D}{\Omega}_0,\\
        \qquad\boldsymbol{\xi}_{\Bar{J}}\cdot\bm{n}_0 = {\Bar{\omega}}_{\Bar{J}} \quad \text{on} \quad \partial_{N}{\Omega}_0.
        \end{split}
    \end{equation}
\end{itemize}
The prescribed microstress couple is set to ${\Bar{\omega}}_{\Bar{J}}=0$ for the remainder of this study, as we do not postulate any external action on the boundary at the microscopic level.

The Helmholtz free energy depends on both external and internal state variables and can be written as $\psi(\mathbf{F}, \Bar{J}, \nabla\Bar{J})$. Using this representation, its material time derivative is obtained via the chain rule as
\begin{equation}\label{rateFreeEnergy}
    \frac{\textnormal{d}\psi}{\textnormal{d} t} = \frac{\partial\psi}{\partial\mathbf{F}}\colon\dot{\mathbf{F}} + \frac{\partial \psi}{\partial \Bar{J}}\dot{\Bar{J}} + \frac{\partial \psi}{\partial \nabla\Bar{J}}\cdot\nabla\dot{\Bar{J}}.
\end{equation}

To satisfy the second law of thermodynamics, we apply the Coleman–Noll procedure and combine it with the internal power expression in Eq. \eqref{rateIntPower}, resulting in the following expression for the energy dissipation rate:
\begin{equation}\label{rateDis}
   \mathcal{D} = \left(\mathbf{P} - \frac{\partial\psi}{\partial\mathbf{F}}\right)\colon\dot{\mathbf{F}} + \left(f_{\Bar{J}} - \frac{\partial \psi}{\partial \Bar{J}}\right)\dot{\Bar{J}} + \left(\boldsymbol{\xi}_{\Bar{J}} - \frac{\partial \psi}{\partial \nabla\Bar{J}}\right)\cdot\nabla\dot{\Bar{J}} \geq 0.
\end{equation}

\subsubsection{Constitutive relations}
Requiring the inequality in Eq.\eqref{rateDis} to hold for arbitrary rates of the state variables leads to the following constitutive relations:
\begin{equation}\label{constitutive}
    \mathbf{P} = \frac{\partial\psi}{\partial\mathbf{F}}, \quad
    f_{\Bar{J}} = \frac{\partial \psi}{\partial \Bar{J}}, \quad
    \boldsymbol{\xi}_{\Bar{J}} = \frac{\partial \psi}{\partial \nabla\Bar{J}} 
\end{equation}
which can be used for the Helmholtz free energy densities presented in Eqs.\eqref{psi-iso-VW3}-\eqref{psi-iso-VW4}, leading to
\begin{equation}\label{constitutives}
\begin{split}
    \boldsymbol{\xi}_{\Bar{J}} = \ell^2\nabla\Bar{J}\quad \text{in} \quad \Omega_0, \\
    f_{\Bar{J}} =  -\frac{nk_{b}\theta}{\chi}\left[\frac{1}{\Bar{J}-f_0}-\frac{\epsilon_{a}}{k_{b}T}\frac{f_{0}}{\Bar{J}^2}\right] -c \left[J - \Bar{J}\right] \quad \text{in} \quad \Omega_0,
\end{split}
\end{equation}
noting that the first Piola-Kirchhoff stress tensor will have different expressions according to the choice from  Eqs.\eqref{psi-iso-VW3}-\eqref{psi-iso-VW4}.

\subsection{Strong form} \label{section:strong_weak_forms}
Incorporating the constitutive laws into the governing equations in Eqs. \eqref{mechanical_equilibrium_governing} and \eqref{chain_governing}, we obtain the following set of strong form equations:
\begin{equation}
\begin{split}\label{strong_forms}
    \nabla\cdot\mathbf{P} = \mathbf{0} \quad \text{in} \quad \Omega_0, \\
    \frac{nk_{b}\theta}{\chi}\left[\frac{1}{\Bar{J}-f_0}-\frac{\epsilon_{a}}{k_{b}T}\frac{f_{0}}{\Bar{J}^2}\right] + c \left[J - \Bar{J}\right] + \nabla\cdot\left(\ell^2 \nabla\Bar{J}\right) = 0 \quad \text{in} \quad \Omega_0,
\end{split}
\end{equation}
with the corresponding boundary conditions:
\begin{equation}\label{BC}
\begin{split}
\bm{u} = \Bar{\bm{u}} \quad \text{on} \quad \partial_{D}{\Omega}_0, \\
\mathbf{P}\cdot\bm{n}_0 = \bm{T} \quad \text{on} \quad \partial_N\Omega_0, \\
\Bar{J}= \Bar{\Bar{J}} \quad \text{on} \quad \partial_{D}{\Omega}_0, \\
\nabla\Bar{J}\cdot\bm{n}_0 = 0 \quad \text{on} \quad \partial_{N}{\Omega}_0.
\end{split}
\end{equation}

\subsection{Viscous effects of phase transition}

Ignoring general viscoelastic contributions, which especially for the case of adhesives would have a significant contribution to the response and will be the focus of part III of this series, we focus on introducing viscous dissipation for the anticipated phase transition. As previously discussed, the length scale $\ell$ that has been introduced in this theory relates to network imperfections and ultimately will dictate a nucleus of the phase transition from a dense to rare phase.\footnote{This is the key difference of this theory to the approach from Ball \cite{ball1982discontinuous}. Following the approach presented here, one would anticipate that there would exist no discontinuity of the solution at the origin in the case of studying a sphere under hydrostatic dead load, but a finite region in the material configuration that undergoes a phase transition, and ultimately, the chains in that region lose their load-carrying capacity.} The transition is accommodated by chain disentanglement, which in turn will cause dissipation due to chain-level friction. Ultimately, chains are also expected to rupture, but these effects will be the focus of the study in Part II of this series.  

Based on the first law of thermodynamics, and invoking the Clausius-Duhem inequality, we obtain the dissipation inequality, as expressed in Eq. \eqref{rateDis}. Further, the generalized forces can be additively decomposed into equilibrium and viscous parts
\begin{equation}\label{eq::viscousdecoupling1}
    \mathbf{P} = \mathbf{P}^{\text{eq}} + \mathbf{P}^{\text{visc}}, \quad f_{\Bar{J}} = f_{\Bar{J}}^{\text{eq}} + f_{\Bar{J}}^{\text{visc}}, \quad \boldsymbol{\xi}_{\Bar{J}} = \boldsymbol{\xi}_{\Bar{J}}^{\text{eq}} + \boldsymbol{\xi}_{\Bar{J}}^{\text{visc}} \, .
\end{equation}
As the aim here is not to consider general viscoelastic effects, we confine the dissipative effects to the conjugate force of the nonlocal volumetric deformation $f_{\Bar{J}}^{\text{visc}}$, that will be triggered from a phase change and therefore set 
\begin{equation}\label{eq::viscousdecoupling2}
    \mathbf{P}^{\text{visc}} = 0, \quad \boldsymbol{\xi}_{\Bar{J}}^{\text{visc}} = 0 \, .
\end{equation}
It is noted that in polymer mechanics, chain-level friction is usually captured through following the deviatoric part of the deformation, whereas here extreme volumetric deformation rates are expected to contribute to chain disentanglement. The equilibrium parts follow directly from the free energy density,
\begin{equation}\label{eq::eqmgeneralized}
    \mathbf{P}^{\text{eq}} = \frac{\partial \psi}{\partial \mathbf{F}}, \quad f_{\Bar{J}}^{\text{eq}} = \frac{\partial\psi}{\partial\Bar{J}}, \quad \boldsymbol{\xi}_{\Bar{J}}^{\text{eq}} = \frac{\partial\psi}{\partial\nabla\Bar{J}} \, .
\end{equation}
Substituting Eq. (\ref{eq::viscousdecoupling1},\ref{eq::viscousdecoupling2},\ref{eq::eqmgeneralized}) in the inequality, Eq. \eqref{rateDis} yields
\begin{equation}
    \mathcal{D} = f_{\Bar{J}}^{\text{visc}}\dot{\Bar{J}} \, .
\end{equation}
To ensure that $\mathcal{D}\geq0$ for arbitrary rates $\dot{\Bar{J}}$, following Gurtin, we introduce a dissipation potential \cite{gurtin1996generalized}, 
\begin{equation}
    \mathcal{R}\left(\Bar{J}\right) = \frac{1}{2}\eta\dot{\Bar{J}}^2, \quad \forall \quad \eta \geq 0
\end{equation}
where $\eta$ is the viscosity of the phase transition\footnote{Noting that similar viscous penalization is often referred to as artificial viscosity in the context of numerical schemes such as phase field methods\cite{miehe2010rate}.}, leading to the viscous microforces of generalized standard material form:
\begin{equation}
    f_{\Bar{J}}^{\text{visc}} = \frac{\partial\mathcal{R}}{\partial\Bar{J}} = \eta\dot{\Bar{J}} \, .
\end{equation}
The local dissipation density is therefore
\begin{equation}
    \mathcal{D} = \eta\dot{\Bar{J}}^2 \geq 0
\end{equation}
guaranteeing thermodynamic admissibility.
Revisiting the microforce balance leads to 
\begin{equation}
    \frac{\partial\psi}{\partial\Bar{J}} + \nabla\cdot\left(\frac{\partial\psi}{\partial\nabla\Bar{J}}\right) + \eta\dot{\Bar{J}} = 0
\end{equation}
which eventually takes the form:
\begin{equation}
\label{strong_forms_viscosity}
    \eta\dot{\Bar{J}} + \frac{nk_{b}\theta}{\chi}\left[\frac{1}{\Bar{J}-f_0}-\frac{\epsilon_{a}}{k_{b}T}\frac{f_{0}}{\Bar{J}^2}\right] + c \left[J - \Bar{J}\right] + \nabla\cdot\left(\ell^2 \nabla\Bar{J}\right) = 0 \quad \text{in} \quad \Omega_0,
\end{equation}
that can be directly substituted in the strong form in Eq.\eqref{strong_forms}. It is noted that the form here acts globally and not just on regions that have undergone a phase transition. The logic behind this formulation is the following: as discussed earlier, the fluid-like contribution to the free energy maintains near incompressibility until a threshold is met and a phase transition takes place, allowing the volume ratio to grow in an unstable fashion, enabling high rates of volumetric deformation. Regions that have not undergone a phase transition will maintain a local and nonlocal volume ratio $\approx1$. As such, the model for viscous contributions selected here is expected to have a substantial contribution in regions that have undergone a phase transition. Physically meaningful selection of $\eta$ is expected not to show significant viscous effects in the absence of phase transitions.

\subsection{Finite element implementation}\label{Section:NumericalImplementation}

To discretize the governing weak form, a mixed finite element approach is employed. The displacement field $\bm{u}$ is approximated using second-order continuous Lagrange elements ($\mathbb{P}_2$), while the scalar internal variable $\Bar{J}$ is represented with first-order Lagrange elements ($\mathbb{P}_1$).
The mixed function space is then defined as:
\begin{align*} \label{trial_spaces}
\mathbb{T}_{trial} &= \mathbb{U} \times \mathbb{J}, \\
\mathbb{U} &=\{ \mathbf{u} \in [H^1(\Omega_0)]^d; \bm{u} = \Bar{\bm{u}} \text{ on } \partial_D\Omega_0\}, \\
\mathbb{J} &=\{ \Bar{J} \in H^1(\Omega_0)\},
\end{align*}
where $H^1$ is the first order Sobolev space and $d \in \{2, 3\}$ denotes the spatial dimension of the problem. Correspondingly, we define the test functions $(\bm{v}, \beta) \in (\mathbb{V}, \mathbb{B})$ within the following spaces:
\begin{align*}
\mathbb{T}_{test} &= \mathbb{V} \times \mathbb{B}, \\
\mathbb{V} &= \{ \mathbf{v} \in [H^1(\Omega_0)]^d; \bm{v} = \bm{0} \text{ on } \partial_D\Omega_0\}, \\
\mathbb{B} &= \left\{ \beta \in H^1(\Omega_0) \right\}.
\end{align*}
By applying the standard Galerkin procedure, we derive the following weak form (considering $\nabla\Bar{J}.\mathbf{n}=0$):
\begin{equation}\label{weak_form_GED}
\begin{split}
    \int_{\Omega_0} \mathbf{P} \colon \nabla\bm{v} \,dV - \int_{\partial_N\Omega_0} \Bar{\bm{T}} \cdot \bm{v} \,dA & = 0, \\
\int_{\Omega} \eta \, \dot{\bar{J}} \, \beta \, d\Omega
+ \int_{\Omega} \frac{n k_{B} \theta}{\chi} 
\left[ \frac{1}{\bar{J} - f_{0}} 
      - \frac{\varepsilon_{a}}{k_{B} T} \frac{f_{0}}{\bar{J}^{2}} 
\right] \beta \, d\Omega
+ \int_{\Omega} c (J - \bar{J}) \, \beta \, d\Omega
- \int_{\Omega} \ell^{2} \nabla \bar{J} \cdot \nabla \beta \, d\Omega & = 0.
\end{split}
\end{equation}

The cavitation model is implemented in the \texttt{FEniCS} finite element framework \cite{alnaes2015fenics}, utilizing the automatic differentiation capabilities of the Unified Form Language (UFL). The full Python implementation is openly accessible on GitHub\footnote{\url{https://github.com/MMousavi98/cavitation}}. The displacement field $\bm{u}$ and the nonlocal volume ratio field $\Bar{J}$ are discretized over the domain as:

\begin{equation}\label{eq_displacement}
\bm{u}(\bm{x})= \sum_{i=1}^{N} N_i(\bm{x}) \bm{u}_i, \quad
\Bar{J}(\bm{x})= \sum_{i=1}^{N} N_i(\bm{x}) \Bar{J}_i
\end{equation}
where $N_i(\bm{x})$ are the basis (shape) functions associated with node $i$, and $\bm{u}_i$ and $\Bar{J}_i$ denote the nodal degrees of freedom for displacement and nonlocal volume ratio, respectively. As mentioned earlier, to maintain stability and prevent pressure oscillations in the mixed formulation, we adopt a Taylor–Hood element pairing, using quadratic interpolation for $\bm{u}$ and linear interpolation for $\Bar{J}$.

The model assumes quasi-static loading and is solved incrementally using a load-ramping procedure. At each load step, the coupled system, consisting of the mechanical equilibrium and nonlocal volume ratio equations (Eqs. \eqref{weak_form_GED}$_1$ and \eqref{weak_form_GED}$_2$), is solved using the built-in \texttt{SNES} nonlinear solver with the \texttt{newtonls} method. A backtracking line search (\texttt{bt}) is employed with convergence tolerances of $10^{-6}$ for absolute, relative, and solution norms, and a maximum iteration limit of $200$.
\section{Results and Discussion}\label{Section:results}
\begin{figure}[h!]
    \centering
    \includegraphics[width=0.7\linewidth]{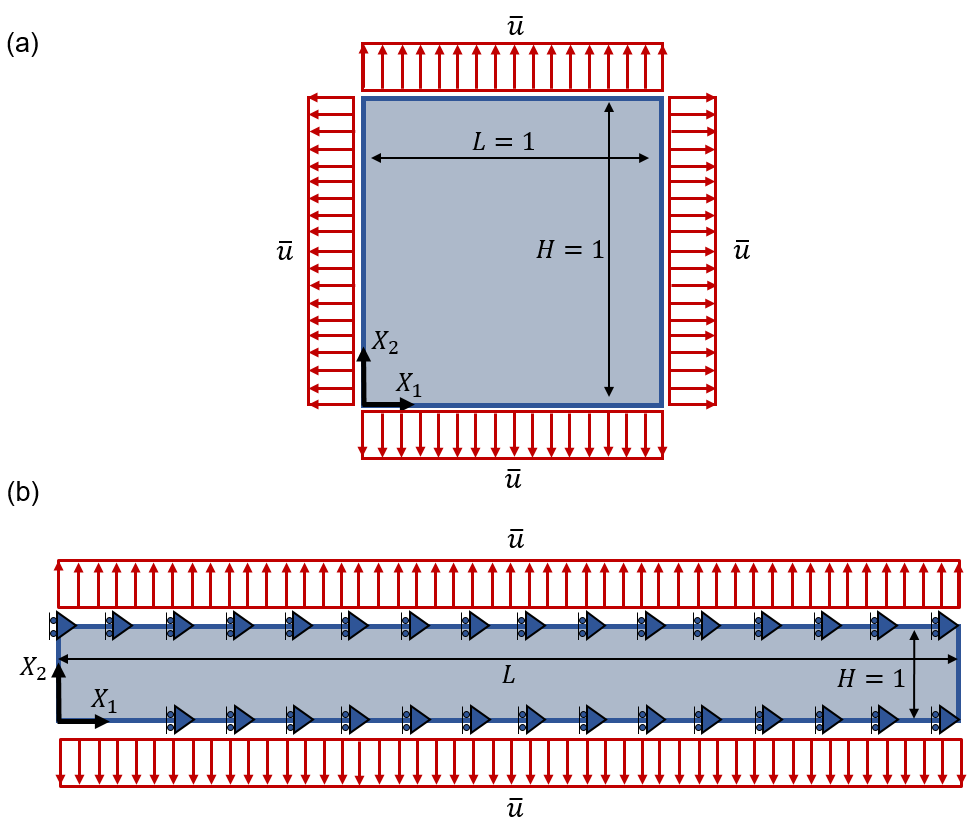}
    \caption{Schematic of the Boundary Value Problems (BVPs) and loading conditions. (a) A constrained biaxial test under plane strain conditions. (b) A ``pure shear" test under plane strain conditions.}
    \label{fig:BVPs}
\end{figure}

In this section, we demonstrate the performance of the proposed continuum cavitation model through two distinct boundary value problems (BVPs), (the later involving a geometric parametrization) and by employing two different free energy formulations; namely, the Neo-Hookean model and the monodisperse network model, as examples of strain softening and strain stiffening models which are two classes that have shown to have very distinct responses in the study of cavitation. In all simulations, the Helmholtz free energy function $\psi$ and the first Piola–Kirchhoff stress $\mathbf{P}$ are normalized by the shear modulus $\mu$. Moreover, all spatial coordinates and displacements are non-dimensionalized using the characteristic domain length $H$ (illustrated in Fig. \ref{fig:BVPs}) for each problem.

In the first BVP, we consider a square domain of dimensions $L \times H = 1 \times 1$, analyzed under plane strain conditions and subjected to constrained bi-axial tension, examining a 2D variant of the cavitation problem, noting that this induces a tri-axial but not hydrostatic state. Normal components of displacement at domain boundaries are ramped up to  $\Bar{\mathbf{u}}\cdot \mathbf{n}_0= 0.2$, as shown in Fig. \ref{fig:BVPs}(a). Each boundary is free to move tangentially, while the prescribed displacements are imposed only in the normal direction. The displacement loading is applied incrementally over 200 steps \footnote{For certain studies, this may differ and will be stated explicitly.}. For the nonlocal field $\Bar{J}$, homogeneous Neumann boundary conditions are imposed along all boundaries, i.e., $\nabla \Bar{J} \cdot \bm{n}_0 = 0$. This problem is first solved using the Neo-Hookean free energy (Eq. \eqref{psi-iso-VW1}). To incorporate more detailed chain statistics, the Neo-Hookean model is subsequently replaced by the monodisperse network formulation (Eq. \eqref{psi-iso-VW2}).

In the second BVP, we analyze a ``pure-shear" test in plane strain conditions, utilizing a rectangular domain of size $L \times H$, as illustrated in Fig. \ref{fig:BVPs}(b). A prescribed displacement is ramped up to $\Bar{u}_2= \pm 0.2$, applied to the top and bottom boundary in the $X_2$ direction, such that the specimen is in tensile loading. The displacement in the $X_1$ direction is $\Bar{u}_1=0$ at the top and bottom boundaries, and the left and right boundaries are traction-free. As in the first problem, each boundary is free to move tangentially, and displacements are applied only in the normal direction. The loading is incrementally applied over 600 steps. The nonlocal field $\Bar{J}$ again satisfies homogeneous Neumann boundary conditions across all boundaries, i.e., $\nabla \Bar{J} \cdot \bm{n}_0 = 0$. All simulations for this BVP are conducted using the monodisperse network model.

For both the Neo Hookean and the statistical mechanics-based constitutive models, stresses are normalized with the shear modulus $\mu=nk_{b}\theta$. Unless noted otherwise, all simulations use $\mu = 1$, $X = 0.2$, $f_0 = 0.85$, $\epsilon_a = 10$, $c=100$, and $\eta = 20$ (time steps are taken at normalized time increments of $\Delta t=1$). The bond stiffness is $E = 1000$, and the monodisperse network model has $N = 5$ Kuhn segments, except for specific special cases. Additionally, the reference configuration $\Omega_0$ (where $\mathbf{F}=\mathbf{I}$) is not assumed to be a stress-free state. The initial state that corresponds to the stress-free equilibrium configuration is determined by finding the isotropic stretch, $\lambda_0$, such that $\mathbf{F}_0=\lambda_0\mathbf{I}$, that satisfies this condition. This is done by assuming a pure volumetric deformation $\mathbf{F} = \lambda_0 \mathbf{I}$, and then solving the resulting nonlinear equation $\mathbf{P}(\lambda_0 \mathbf{I}) = \mathbf{0}$ for $\lambda_0$. To begin the simulation from this stable state, an initial displacement field $\bm{u}(\bm{X}) = (\lambda_0 - 1)\bm{X}$ is prescribed, which maps the reference coordinates to the initial stress-free configuration. Finally, the length scale $\ell$ is set to $0.05$ for simulations using the Neo-Hookean model and $0.15$ for those using the monodisperse network model unless noted otherwise. The results section is organized into five parts. The first subsection presents the simulation of cavitation in the first BVP using the Neo-Hookean model. Two parametric studies are also conducted to investigate the effects of mesh refinement and viscosity of phase transition. In the second subsection, we address the broadening of the cavity observed in the reference configuration and propose a suitable modification to obtain physically meaningful results. The third subsection replaces the Neo-Hookean formulation with the monodisperse network model and re-examines the first BVP under this free energy function. In the fourth part, we focus on the second BVP and analyze the influence of mesh density and length scale on the results. Finally, the fifth subsection explores the effect of the domain aspect ratio on cavitation behavior.

\subsection{Constrained biaxial tension: Neo-Hookean model}

We begin by analyzing the first BVP using the Neo-Hookean free energy model. This serves as the baseline case to demonstrate the model's ability to capture cavitation. As the solution prior to a phase transition is expected to be homogeneous, we introduce a small material imperfection at the center of the domain. This is implemented by reducing the shear modulus $\mu$ by $1\%$ in a circular disc of radius $R=0.01$ with its origin at the center of the rectangular domain. This wouldn't be necessary if we had implemented periodic boundary conditions for the specific problem. It is noted that for solutions that are heterogeneous prior to the phase transition, this will not be necessary, as demonstrated later with the second (more realistic) BVP. 

\begin{figure}[h!]
    \centering
    \includegraphics[width=0.8\linewidth]{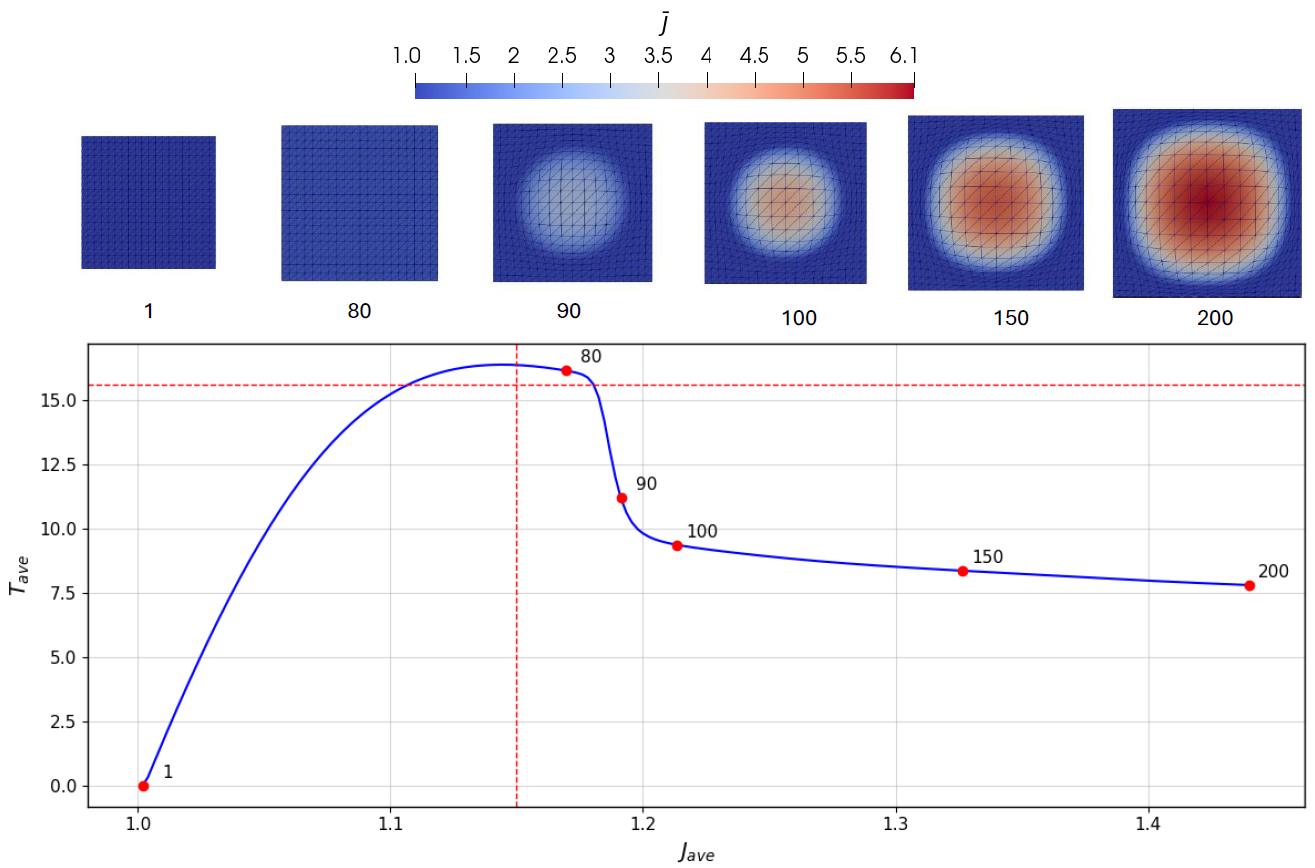}
    \caption{Macroscopic mechanical response of the square domain using the Neo-Hookean model. The red dashed line marks the analytical threshold ($J_{critical} \approx 1.15$) where the free energy loses convexity with respect to $J$. The red dots correspond to the field snapshots shown on top of the graph for steps 1, 80, 90, 100, 150, and 200.}
    \label{fig:first_case_graph}
\end{figure}

The overall mechanical response is presented in Fig. \ref{fig:first_case_graph}, which plots the average current normal traction $T_{ave}$\footnote{The average normal traction (in the current configuration, a spatial quantity) is computed via the boundary integral of the First Piola-Kirchhoff stress tensor normalized by the current deformed width: $T_{ave} = \frac{1}{L_{curr}} \int_{\Gamma_{top}} (\mathbf{P} \cdot \mathbf{N}) \cdot \mathbf{e}_y \, dS$.} against the average volume ratio $J_{ave}$ of the domain, to mirror a pressure vs. volume change plot that would be relevant in a 3D setting  (here it is noted that the corresponding triaxial state is not hydrostatic). The material initially shows a monotonic response, with the traction reaching a peak value and continuing at a mild descending plateau. Immediately following this plateau, the curve exhibits a sharp, sudden decrease in the load-carrying capacity (around $40\%$), indicating an unstable event. This drop corresponds to the onset of cavitation. This delay is indicative of the nonlocal features that we have added to the theory, as now sharp discontinuities are not allowed, and additional energy is required to form the diffuse interfaces between the rare and dense phases. Fig. \ref{fig:first_case_graph} also provides the corresponding contour plots of the nonlocal volume ratio, $\Bar{J}$, at key load steps in the current configuration. At step 1, the material is in its reference state ($\Bar{J}=1$). By step 80, corresponding to the end of the plateau, the volumetric deformation is still largely homogeneous (noting that there is a small heterogeneity in the field due to the introduced material imperfection). At step 90, towards the end of the unstable event, a distinct region of high volumetric expansion ($\Bar{J} > 3$) nucleates at the center, confirming the formation of a region that has undergone a phase transition. We will refer to this region as a "cavity", even though we note that the material in this region is still virtually undamaged and can resist shear, but has effectively lost resistance to bulk deformations corresponding to the van der Waals free energy contribution\footnote{The phase transition is reversible at this stage as no irreversibility conditions have been enforced.}. Simultaneously, the dense region that surrounds the cavity maintains a near-incompressible state with low volumetric deformations $\Bar{J}\approx1$. We note that the dynamics of the phase transition are due to the dissipation effects that we have incorporated in the model. In the subsequent steps (100, 150, and 200), this internal cavity clearly expands in the current configurations, while the traction required for continued expansion settles at a lower value. The results show a clear correlation: the sudden, unstable drop in traction seen in Fig. \ref{fig:first_case_graph} is a direct macroscopic consequence of the microscopic cavity nucleation and growth visualized in the contours.

It is worth noting that the onset of this instability, driven by the van der Waals component of the free energy, can be predicted analytically as discussed in Lamont {\it{et al.}} (2026) \cite{lamont2025cohesive}. For purely volumetric deformations, the instability occurs when the free energy function (Eq. \eqref{psi-iso-VW1}) loses its convexity with respect to $J$, defined by the condition $\partial^2 \psi / \partial J^2 \leq 0$. For more general deformation, the conditions for loss of ellipticity have to be examined. This marks an energetic turning point where the system favors rapid expansion to lower its total elastic energy. By solving for this bifurcation ($\partial^2 \psi / \partial J^2 = 0$) using the material parameters that have been selected (for convenience, the local theory is utilized, without the approximations for the numerical implementation), we find an analytical critical volume change of $J_{critical} \approx 1.15$. This theoretical threshold is depicted by the red dashed line in Fig. \ref{fig:first_case_graph}. The simulation approaches the point of instability that is analytically calculated both regarding the traction and volumetric deformation. 
We explore the effects of both mesh refinement and viscosity of phase transition in the subsequent parametric studies.

\begin{figure}[h!]
    \centering
    \includegraphics[width=1\linewidth]{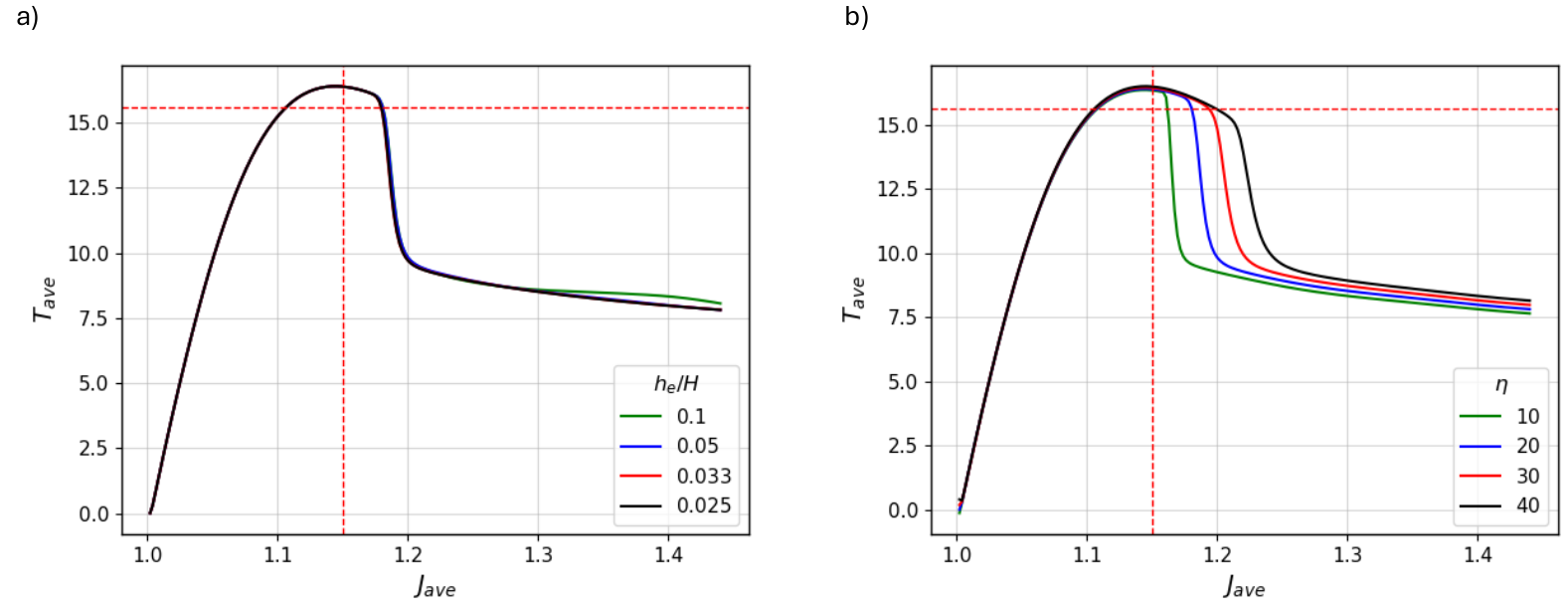}
    \caption{Sensitivity analysis of the Neo-Hookean model. (a) Effect of mesh refinement on the traction-volume response, showing convergence as the element size $h_e/H$ decreases from 0.1 to 0.025. (b) Influence of the viscosity of phase transition ($\eta$) on the onset of cavitation. Lower viscosity values result in a peak traction closer to the analytical instability threshold (red dashed line).}
    \label{fig:first_case_mesh_ell}
\end{figure}

\begin{figure}[h!]
    \centering
    \includegraphics[width=0.7\linewidth]{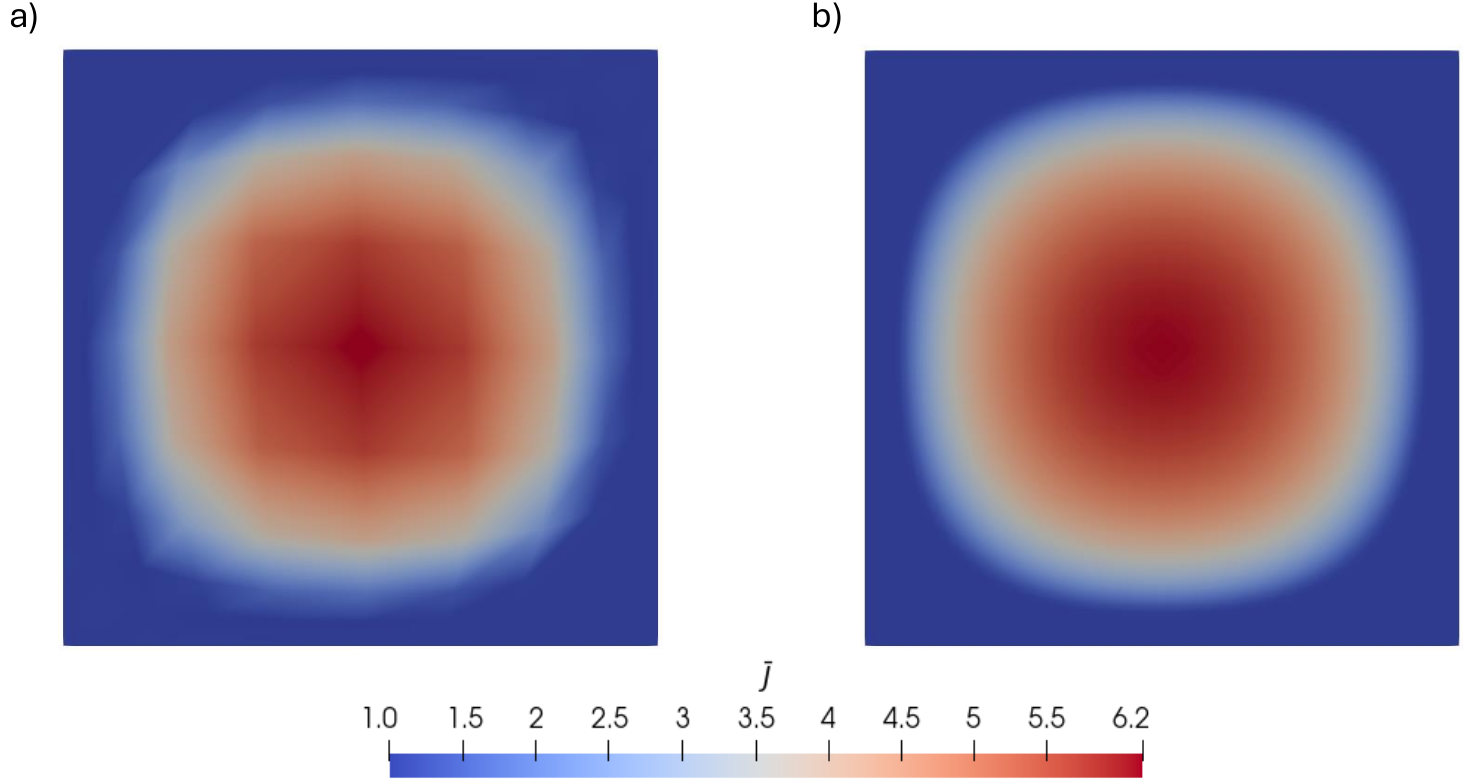}
    \caption{Impact of mesh discretization (a) coarse and b) fine meshing) on the resolution of the cavity field.}
    \label{fig:first_case_meshing}
\end{figure}

We use four different element sizes in the range of $h_e/H=0.1$ to $h_e/H=0.025$ to evaluate the sensitivity of the model to the meshing.  The results are presented in Fig. \ref{fig:first_case_mesh_ell} (a). This plot clearly demonstrates convergence: as the element size decreases, the traction-volume response curves merge. The solutions for $h_e/H = 0.033$ and $h_e/H = 0.025$ are nearly identical, indicating that a sufficiently refined mesh ($h_e/H=0.05$) yields a convergent result both for the principal solution and the post-bifurcation response.  Contour plots for two of the four cases are presented in Fig. \ref{fig:first_case_meshing}, illustrating the deformed mesh in the current configuration with good agreement.  Additionally, the effect of viscosity of phase transition is examined in Fig. \ref{fig:first_case_mesh_ell} (b). As the viscosity of phase transition $\eta$ decreases, the peak of the curve associated with the onset of the cavitation comes closer to the analytical instability point denoted by the red dashed line. This confirms that the viscosity of the phase transition delays the cavitation. It is noted that prior to the phase transition and corresponding cavitation, there is no prominent viscous effect between the four examined cases.

\subsection{Cavity nucleation and expansion}\label{cavity_broadening}
In the context of cavitation in fluids, the cavitation of water has been modeled in a molecular dynamics setting \cite{wang2017cavitation}. Additionally, simulations of boiling in a continuum setting  \cite{liu2015liquid} incorporated a van der Waals energy to simulate liquid to vapor phase transition in the context of a nonlocal model. In fluids, the cavity is expanding in the current configuration, but in solids, the question is more subtle. For a pre-existing cavity in solids, models like \cite{dollhofer2004surface,lopez2009onset,dal2009micro} (and by extension even Ball \cite{ball1982discontinuous} for the singular limit) treat the cavity expansion as an elastic instability; as such, the boundary of the cavity in the reference configuration is fixed during cavity expansion. In \cite{kang2017cavitation}, pre-existing cavity growth and transition to fractures are examined, but also \cite{kumar2020revisiting}, where assumed micro-defects unstably expand, and continuum damage in the form of fracture takes over. Early works, such as \cite{williams1965spherical}, but also more recent such as \cite{hutchens2016elastic,kim2020extreme} also point to the fact that cavities can expand with diffuse damage without localization of damage in a distinct crack front\footnote{Tangential to this topic, recent studies in phase transitions for fluids \cite{bilby1975change}, but also more recently solids \cite{markenscoff2021m}, suggesting transition from a spherical to penny-shape like growth of the region that undergoes a phase transition.}. As such, one would expect that the model developed here forms a cavity in a finite region through a cohesive instability and corresponding phase transition, and further, there is elastic growth of the cavity in the current configuration while its radius in the reference configuration remains fixed. It is noted that, as a damage formulation has not been developed here, the cavity still carries tensile hydrostatic stresses from the Neo-Hookean contribution\footnote{This wouldn't be the case if the Neo-Hookean formulation just considered the deviatoric stresses, but was seen in statistical mechanics based model presented in Section \ref{Section:theory}, volumetric deformations are also responsible for chain stretching.}.

\begin{figure}[h!]
    \centering
    \includegraphics[width=1\linewidth]{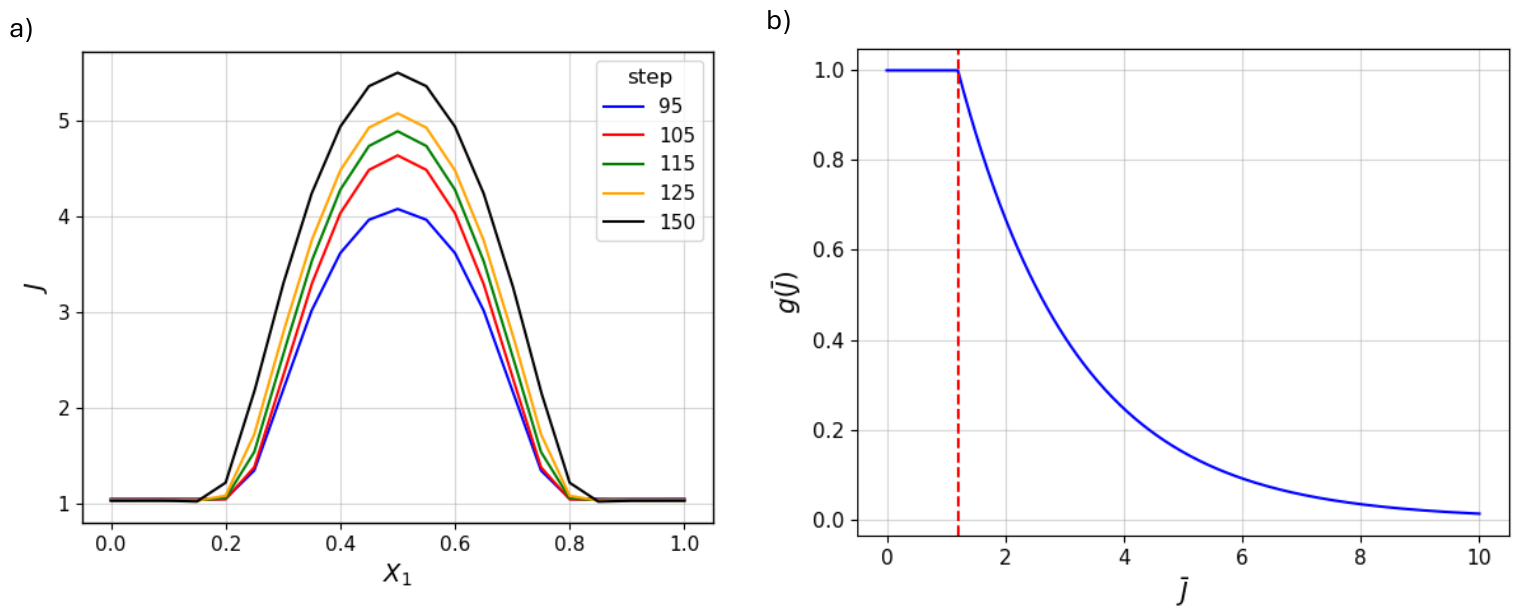}
    \caption{Investigation of numerical cavity broadening and its correction. (a) Cross-sectional profiles of $J$ in the reference configuration at various load steps, illustrating the unphysical broadening inherent in the standard nonlocal formulation. (b) The proposed relaxation function $g(\bar{J})$ decays exponentially after a critical threshold to reduce nonlocal interactions within the cavitated region.}
    \label{fig:broadening}
\end{figure}

While the model successfully captures the onset of cavitation, 
We observe a slight broadening of the cavity's profile in the reference configuration as the deformation progresses. This is shown in Fig. \ref{fig:broadening}\,(a), which plots the volume ratio $J$ at different load steps. 
This broadening is reminiscent of what is seen in gradient damage models \cite{de2016gradient} for which the authors have recently presented a numerical treatment \cite{mousavi2025chain}. Following that treatment, and maintaining thermodynamic consistency, we introduce a relaxation function, $g(\Bar{J})$, which modifies the gradient energy  as
\begin{equation} \label{psi_grd-modified}
    \psi_{grd}(\Bar{J},\nabla\Bar{J}) = \frac{\ell^2}{2} g(\Bar{J}) \nabla\Bar{J} \cdot \nabla\Bar{J}.
\end{equation}
The detailed formulation for this modification is presented in \ref{appendix2}. 

\begin{figure}[h!]
    \centering
    \includegraphics[width=1\linewidth]{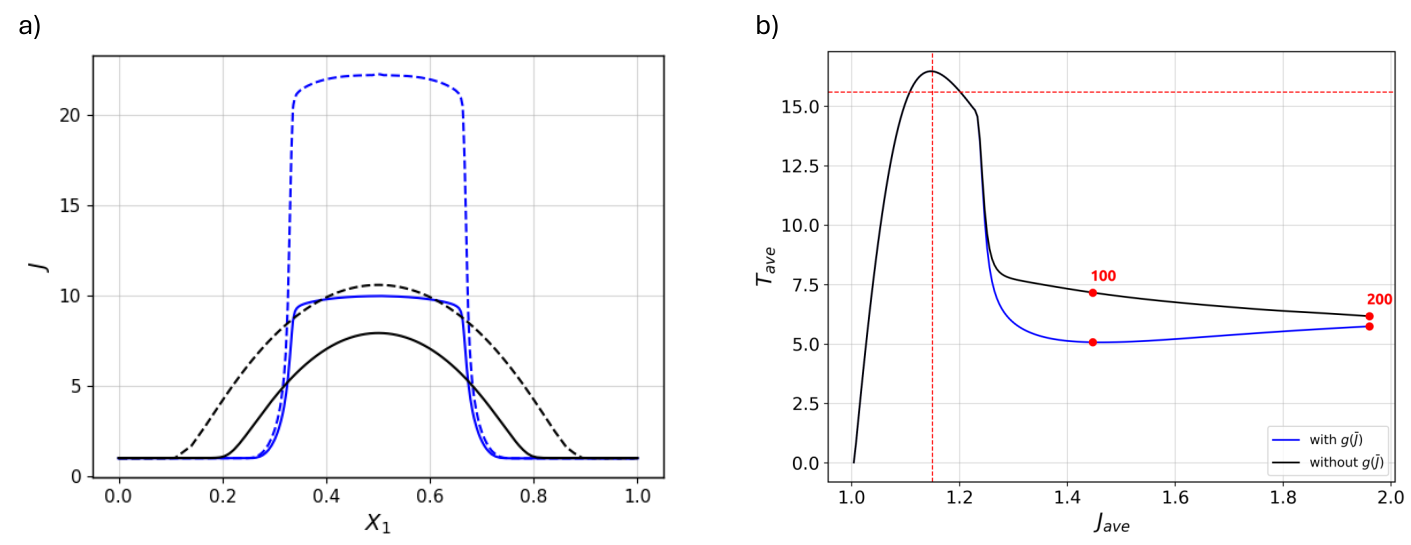}
    \caption{Impact of the relaxation function $g(\bar{J})$ on the simulation results. (a) Comparison of volume change profiles at Step 100 (solid) and Step 200 (dashed), showing that the modified model (blue) eliminates broadening compared to the original model (black). (b) Macroscopic traction-volume response: the modified model exhibits a sharper post-nucleation drop and stabilizes, whereas the original model shows a continuous artificial decline.}
    \label{fig:modified_form}
\end{figure}

The purpose of introducing this function $g(\Bar{J})$ is to progressively diminish the nonlocal interactions once cavitation has initiated \cite{mousavi2025chain}. As shown in Fig. \ref{fig:broadening}\,(b), following \eqref{eq::g-function} $g(\Bar{J})$ is equal to 1 up to the critical threshold $J_c$ and then decays exponentially as $\Bar{J}$ increases. This modification reflects that beyond a certain level, volumetric deformations (which can be very high in the region that has undergone a phase transition) should not further influence the interface between the rare and dense phases. This is especially critical as damage is eventually anticipated (even though we are not treating it here).  The adoption of this relaxation function seemingly alleviates this issue. Fig.~\ref{fig:modified_form}\,(a) contrasts the $J$ profiles (along the $X_1$ direction passing from the center-point of the domain) for the original (black lines) and modified (blue lines) models. The modified model's profile remains stable without broadening. 
It is still an open question, though, how the material length scale $\ell$ relates to the size of the zone that undergoes a phase transition. 

\begin{figure}[h!]
    \centering
    \includegraphics[width=0.7\linewidth]{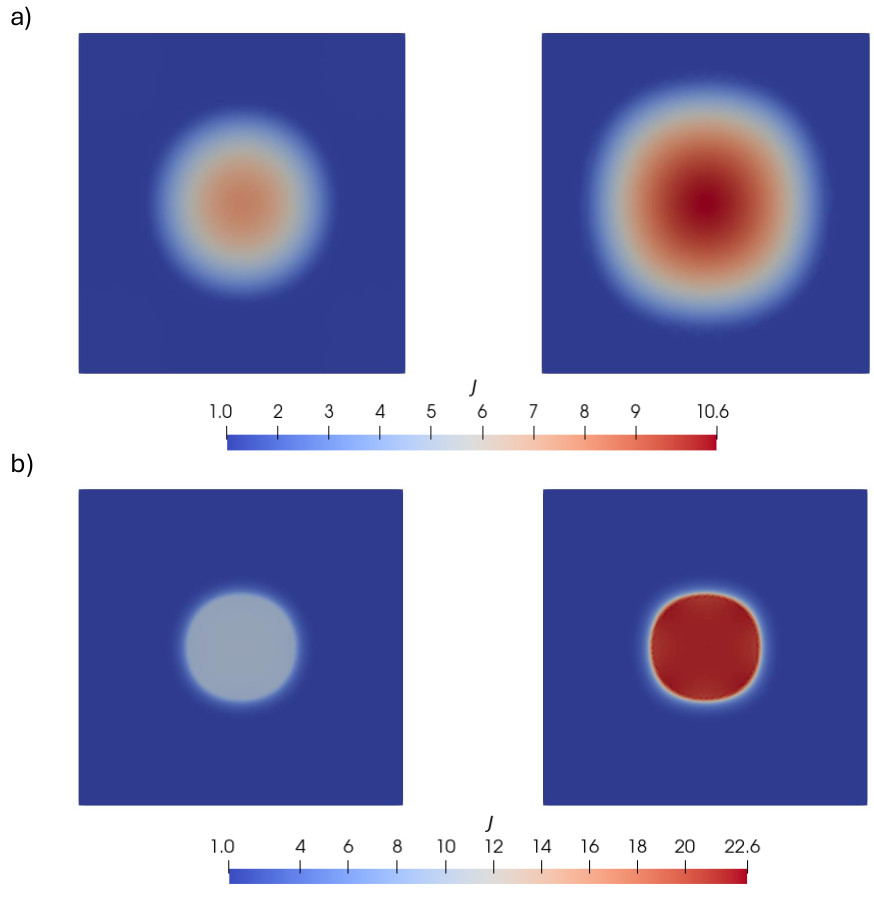}
    \caption{Contour comparison of the cavity field $J$ at steps 100 and 200 in the reference configuration. a) The original model produces diffuse damage zones and a clear broadening. b) The modified model incorporating the relaxation function yields sharp, well-defined cavity interfaces without broadening.}
    \label{fig:modified_form_contours}
\end{figure}

This modification has a direct effect on the macroscopic response as well, especially as it pertains to the post-bifurcation response, as shown in Fig.~\ref{fig:modified_form}\,(b). By adding the relaxation function, a sharper interface emerges, accommodating a higher peak volume change ($\Bar{J}$) and also stabilizing the referential size of the nucleus of the phase transition. The impact of this modification is also captured in Fig.~\ref{fig:modified_form_contours}, which compares the original contours (top row) that significantly expand in the reference configuration at two distinct points in the post-bifurcation response with the sharper and more stable interface produced by the modified model (bottom row).

\begin{figure}[h!]
    \centering
    \includegraphics[width=1\linewidth]{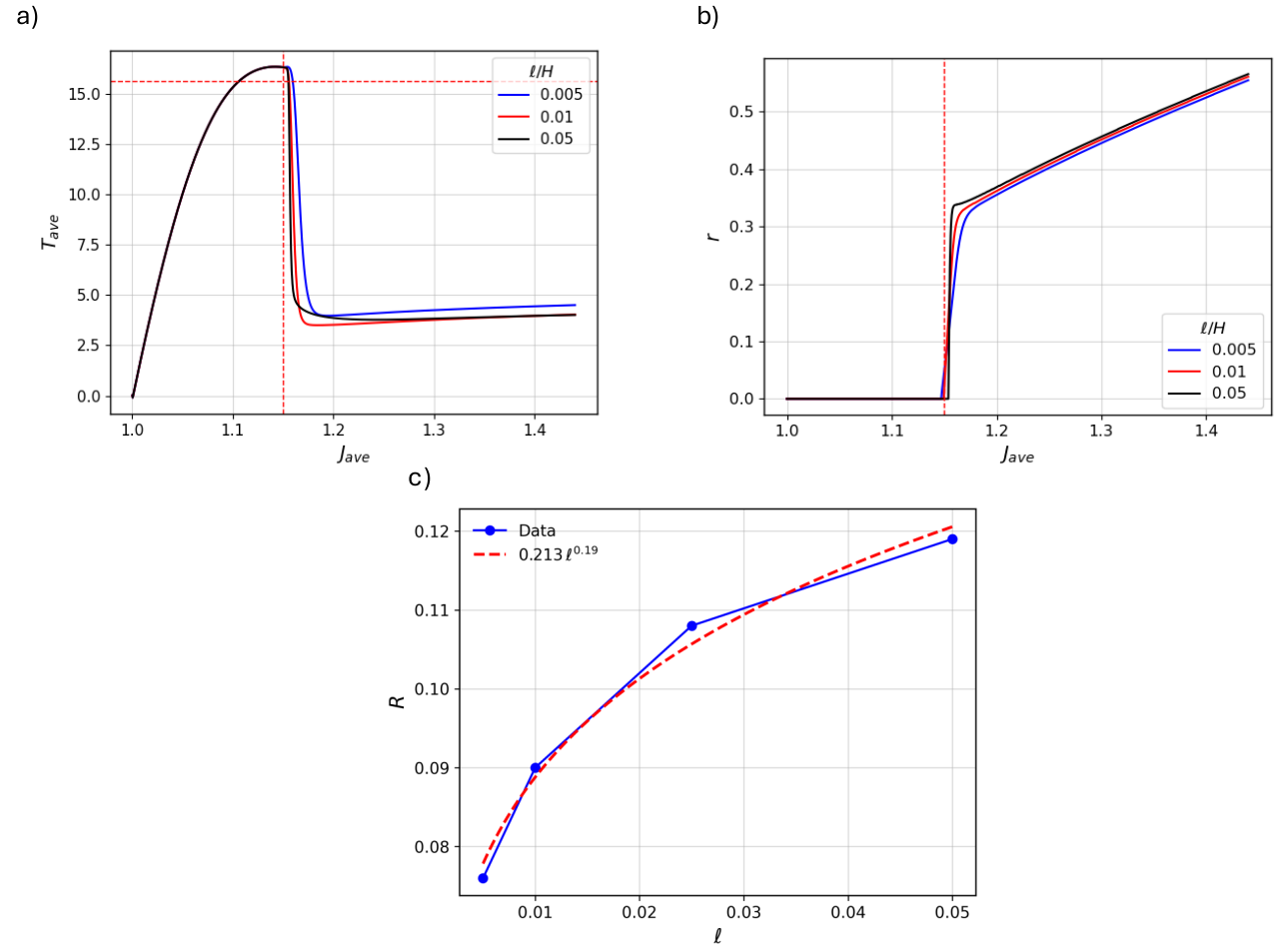}
    \caption{(a) Average traction versus volume change for varying length scales. The nucleation point consistently aligns with the analytical threshold (red dashed line) regardless of $\ell$. (b) Evolution of the cavity radius in the current configuration, showing identical expansion behavior across all length scales. (c) The effect of the length scale on the cavity radius in the reference configuration.}
    \label{fig:pressure_radius_ell}
\end{figure}

We conclude this subsection by examining the influence of the nonlocal length scale, $\ell$. We conduct three simulations with the modified model, using $\ell/H$ values of 0.005, 0.01, and 0.05. We examine these at a slower load ramping rate, getting to the maximum displacement load in 1200 load increments (equivalent to reducing the viscosity of the phase transition).
Furthermore, to properly resolve the smallest length scale, a fine mesh with an element size of $h_e/H = 0.002$ is used for all three cases. The results are presented in Fig.~\ref{fig:pressure_radius_ell}. Several key observations can be made. First, for all three cases, the numerically computed instability threshold approaches the analytical prediction $J_c$ shown in the red dashed line (as expected from the viscosity of phase transition study presented earlier).  Fig.~\ref{fig:pressure_radius_ell}\,(a) shows that the post-peak traction drop is nearly identical for $\ell/H = 0.01$ and $0.05$, but the drop for the smallest length scale, $\ell/H = 0.005$, is slightly less severe and exhibits a minor delay due to more pronounced viscous effects.  The area that undergoes a phase transition appears to follow the 2D equivalent of a spherical pattern for cavitation; the radius of the circle that confines this region can be defined in a reference configuration as $R$ and in the current configuration as $r$\footnote{The current cavity radius was estimated by tracking the spatial extent of the highly expanded material region. Specifically, it was defined as the maximum Euclidean distance from the domain center to any material point satisfying a local volume expansion threshold of $J > 1.5$ in the deformed configuration:
\begin{equation}
    r = \max_{\mathbf{x} \in \Omega} \{ \|\mathbf{x} + \mathbf{u}(\mathbf{x}) - \mathbf{x}_c \| : J(\mathbf{x}) > 1.5 \}
\end{equation}
where $\mathbf{x}$ is the reference coordinate, $\mathbf{u}$ is the displacement, and $\mathbf{x}_c$ is the center of the domain. Note that the threshold $1.5$ was chosen sufficiently above the background volumetric deformation to avoid contamination from bulk fluctuations while fully capturing the region that is undergoing a phase transition.}. Most importantly, Fig.~\ref{fig:pressure_radius_ell}\,(b) demonstrates that the unstable cavity growth and subsequent stable cavity expansion are virtually identical across all three cases as observed in the current configuration.
Fig.~\ref{fig:pressure_radius_ell}\,(c) depicts the effect of the length scale, which is chosen as a physically meaningful material parameter, on the referential nucleus radius of the cavity and corresponding phase transition. These results indicate that an increasing length scale $\ell$ points to cases of higher network imperfection leading to a larger referential radius of the cavity, which can be interpreted as a larger region of the network needing to collectively undergo a phase transition to accommodate cavity formation.

\begin{figure}[h!]
    \centering
    \includegraphics[width=1\linewidth]{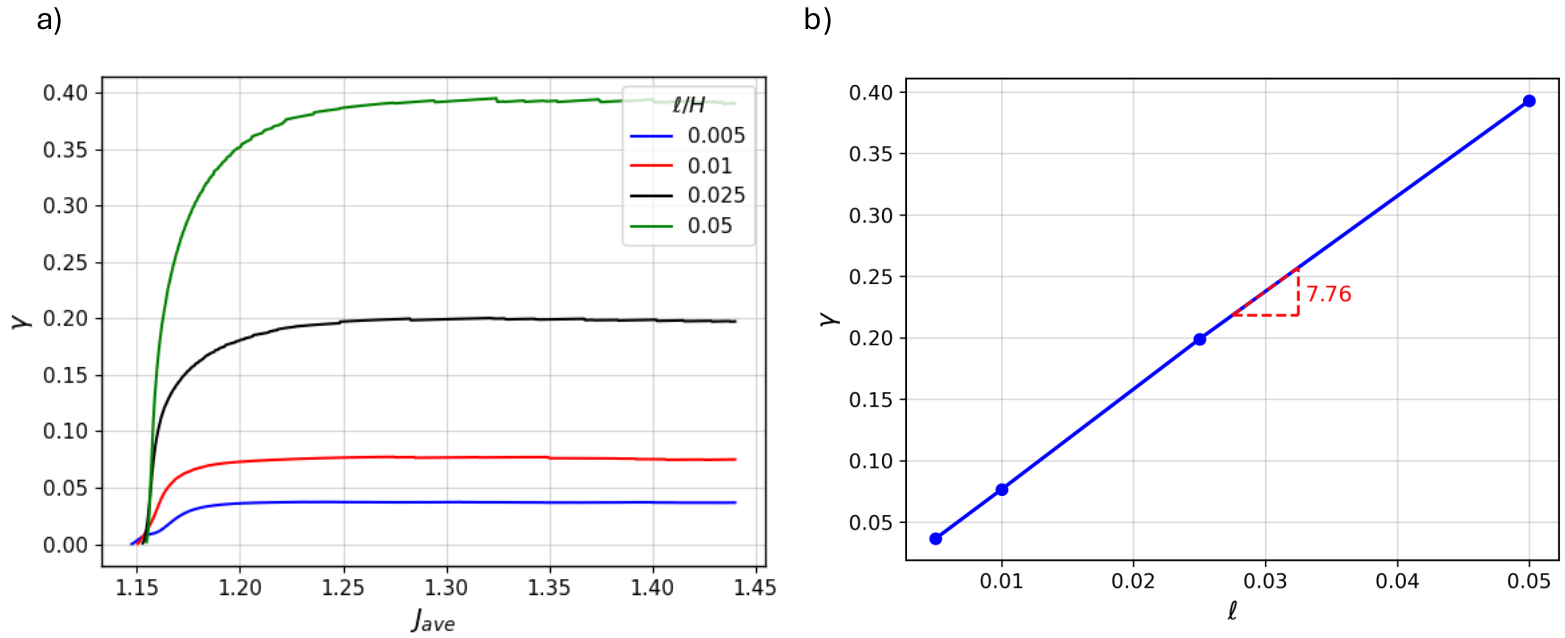}
    \caption{(a) The effect of length scale on the calculated excess interfacial energy per unit referential interface area along the loading path. (b) Exploring the relationship between the length scale and equilibrated excess interfacial energy per unit referential interface area.}
    \label{fig:gamma}
\end{figure}

In Fig. \ref{fig:gamma}, the excess interfacial energy per unit referential interface area is explored. This is numerically evaluated through the corresponding volume integral
\begin{equation}
    \gamma = \int_{\Omega} g(\Bar{J}) \ell^{2} {\nabla \Bar{J}} \cdot \nabla \Bar{J} \, d\Omega /(2\pi R)
\end{equation}
dividing by the perimeter of the cavity in the reference configuration, where $R$ is the detected radius of the cavity (region that has undergone a phase change). Fig. \ref{fig:gamma}(a) shows that the calculated excess interfacial energy in the domain evolving from zero (prior to cavity nucleation) to an equilibrium state captured as the curve plateaus when the cavity is well formed. Fig. \ref{fig:gamma}(b) captures a linear dependence between the equilibrium excess interfacial energy $\gamma$ and the material length scale $\ell$. 

\begin{figure}[h!]
    \centering
    \includegraphics[width=1\linewidth]{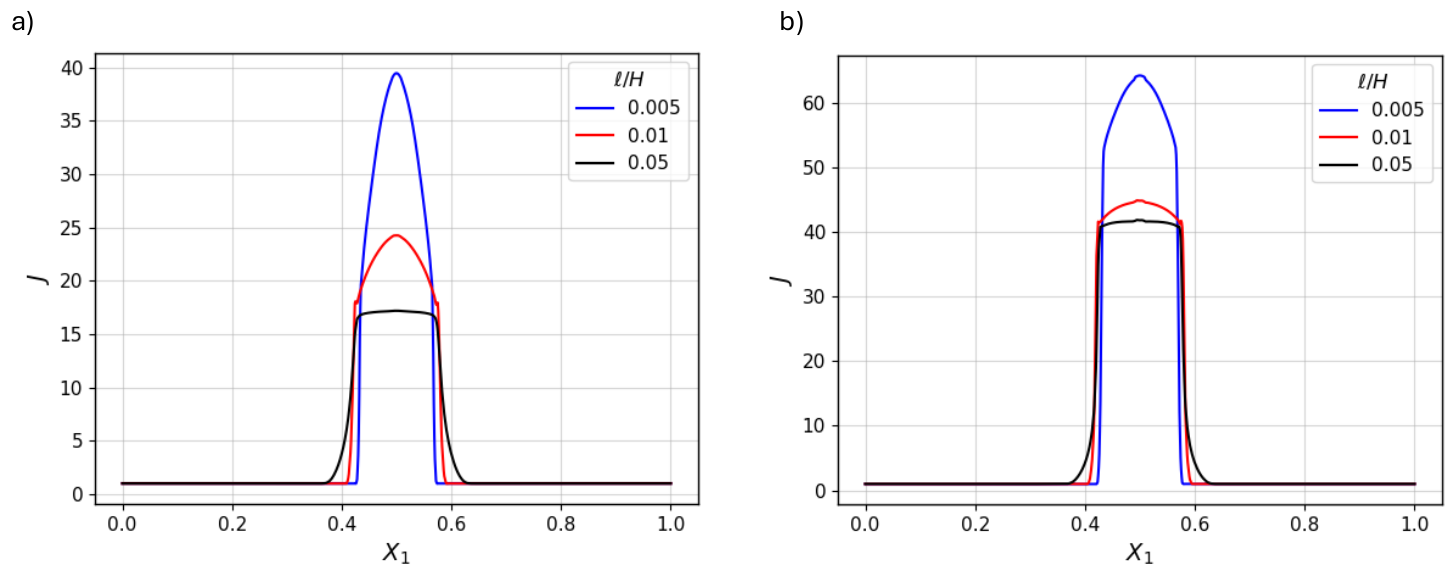}
    \caption{(a) Volume change profiles in the reference configuration at (a) step 600 and (b) 1200 as a function of the length scale $\ell$.}
    \label{fig:reference-radius}
\end{figure}

In Fig. \ref{fig:reference-radius}\,(a) the referential excess interfacial energy is shown to rapidly grow to a constant value for different values of the length scale $\ell$\footnote{A constant referential interfacial surface energy dictates that there are no interfaceial/surface stress effects as discussed in \cite{dollhofer2004surface}. This question will be revisited in part II of this series.}, and Fig. \ref{fig:reference-radius}\,(b) shows that it scales linearly to the material length scale $\ell$. The influence of the length scale $\ell$ is also illustrated in Fig. \ref{fig:reference-radius} where the referential profiles of $J$ are presented at different stages of loading during the post-bifurcation state, once the phase transition has already taken place. These profiles show that decreasing the length scale leads to a smaller cavity radius while producing a markedly larger peak volume change. Viscous effects are also prominent for the smaller $\ell$ cases.

\subsection{Constrained biaxial tension: monodisperse network model}
\begin{figure}[h!]
    \centering
    \includegraphics[width=0.8\linewidth]{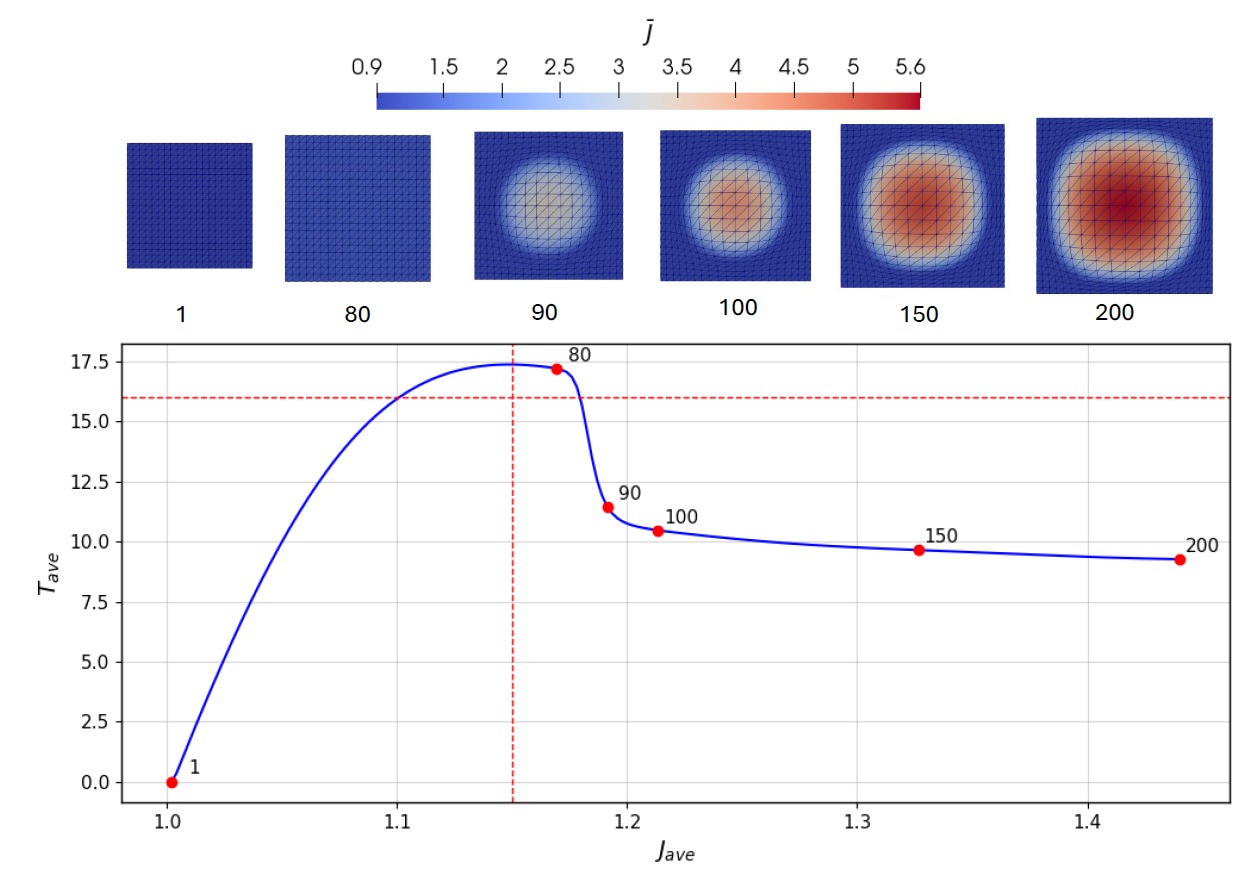}
    \caption{Macroscopic response of the first BVP using the monodisperse network model ($N=5$).}
    \label{fig:second_case_graph}
\end{figure}

A more complex statistical mechanics-based model enables us to incorporate information from the micro- or chain-level scale to the continuum model and examine the potential influence of strain stiffening in the cavitation response. Strain stiffening is present in this model as chains reach their contour length and enthalpic effects dominate.  Note that in the rest of the study, the original formulation proposed for the monodisperse model in Eq. \eqref{psi-iso-StVW5} is utilized. 
The average normal traction to average volume ratio response (following the same convention used previously) and its corresponding $\Bar{J}$ contours for the marked time steps are showcased in Fig. \ref{fig:second_case_graph}. Overall, the trend is very similar to Fig. \ref{fig:first_case_graph} for the Neo-Hookean case, exhibiting the same characteristic traction peak followed by an unstable drop.

\begin{figure}[h!]
    \centering
    \includegraphics[width=0.6\linewidth]{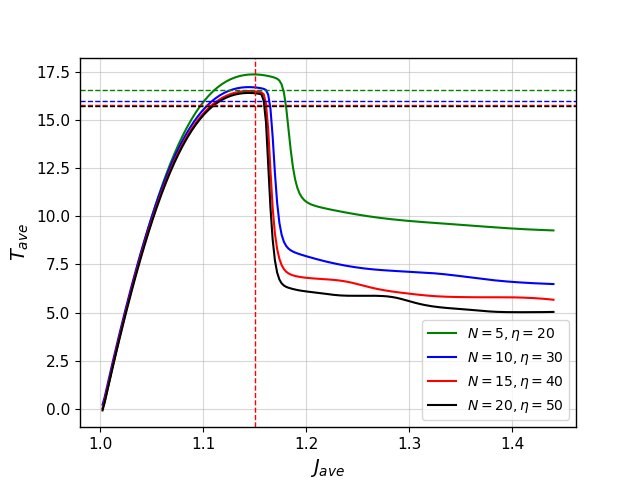}
    \caption{Effect of the number of Kuhn segments ($N$) on the cavitation response. Dashed lines are calculated directly from the instability condition, and solid lines outline the simulation response. Colors for solid and dashed lines are matched.}
    \label{fig:N}
\end{figure}

Fig. \ref{fig:N} illustrates the influence of chain length, controlled by the number of Kuhn segments per chain, $N$, on the cavitation response of the monodisperse network model, evaluated for constant particle density ($nN = \text{const}$, indicating a constant number of Kuhn segments per reference volume). As detailed in \ref{appendix1}, the initial stress-free equilibrium volume ratio ($J_{eq}$) depends on the density ratio $\chi \approx 1/N$. Consequently, the initial referential volumetric deformation varies slightly for each case and is explicitly accounted for in the boundary conditions. Additionally, as the shear modulus scales inversely with $N$ ($\mu \propto n \propto 1/N$)  to maintain constant particle density across the cases that are studied here, the penalty coefficient $c$ is scaled proportionally to maintain a constant $c/\mu$ ratio across all cases. 
The response demonstrates that the onset of the cohesive instability exhibits a weak dependence on the chain length $N$. While decreasing $N$ produces a stiffer overall response (a direct consequence of the higher crosslink density required to maintain the particle density) and higher peak critical load, the corresponding critical volume ratio remains largely unchanged.  At the relatively small stretches corresponding prior to the onset of the instability (e.g., $J_{ave} \approx 1.15$), the polymer chains are far from their finite extensibility limit. In this low-stretch regime, the entropic contribution of the network is not significant. Consequently, the loss of convexity is primarily driven by the van der Waals contribution.  In Fig. \ref{fig:N}, the horizontal dashed lines represent the critical in-plane normal stress, calculated directly from the instability criterion $\partial^2 \psi / \partial J^2 = 0$; the finite element solution is shown with the solid lines. 
Because the van der Waals driving force dominates this early regime, this critical hydrostatic stress remains highly consistent across all examined chain lengths, establishing an intrinsic physical threshold for cavity nucleation that is independent of the background network topology.

\subsection{``Pure shear" test: monodisperse network model}
\begin{figure}[h!]
    \centering
    \includegraphics[width=1\linewidth]{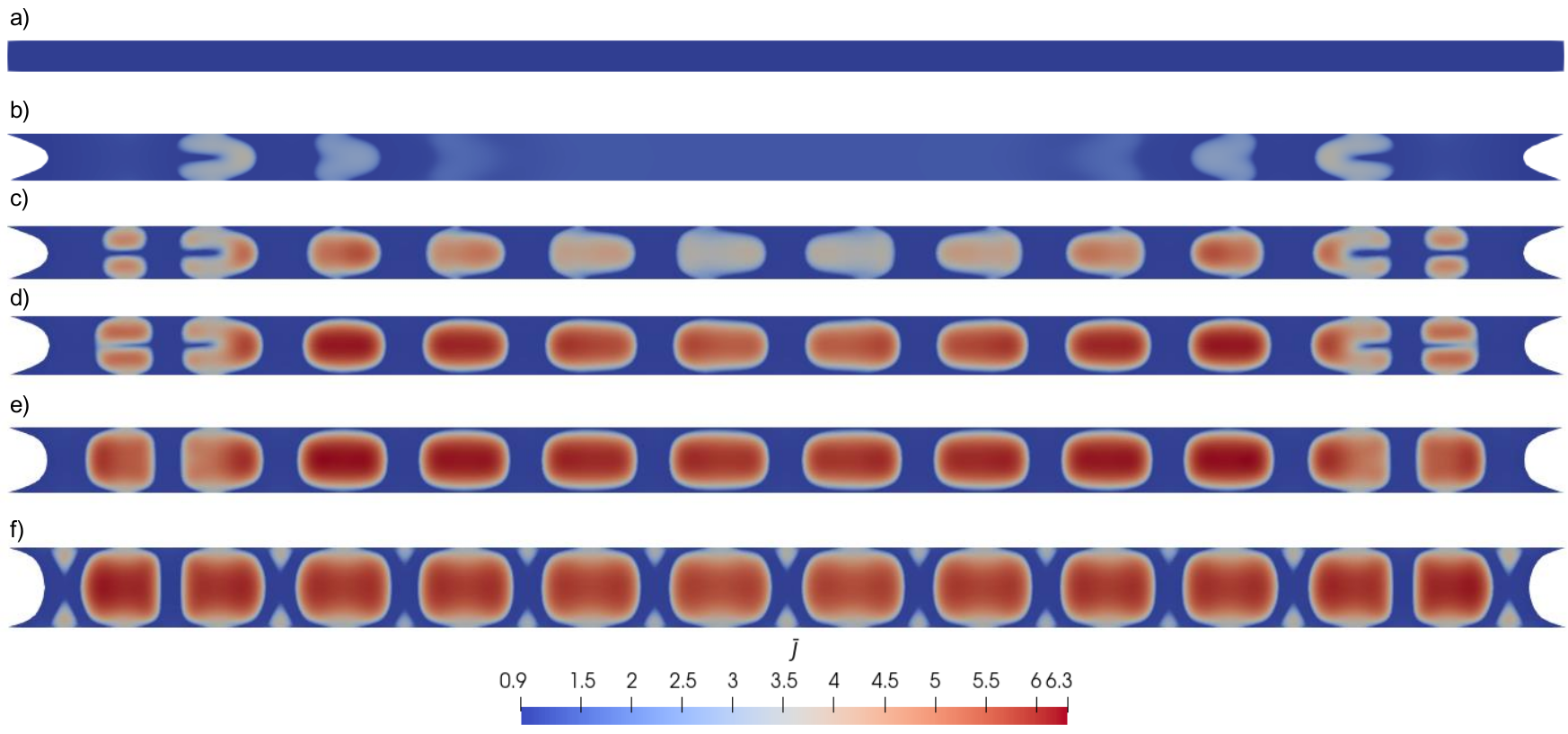}
    \caption{Spontaneous multi-cavity nucleation in the second BVP ($L/H=50$). Evolution of the $\Bar{J}$ field at steps 1, 50, 70, 90, 110, 150, and 200. The model captures a cascade of nucleation events starting from the edges and propagating inward due to the triaxial stress state, without pre-existing defects.}
    \label{fig:strip_contours}
\end{figure}

Having explored the monodisperse network model behavior for a single cavity formation, it is further utilized to examine second BVP (Fig.~\ref{fig:BVPs}\,(b)), a ``pure shear" test often utilized for adhesives and fracture of soft and biological materials. This setup serves as a two-dimensional analogue of the classic "poker-chip" experiment \cite{gent1959internal, guo2023crack}. Relevant for adhesives, this has also recently motivated the study of cavity and crack formation in a thin adhesive film \cite{hao2023does,hao2024constitutive,hao2025approximate}. The displacement is zero in the $X_1$ direction on the top and bottom boundaries.  Crucially, the simulation is initiated on a pristine, homogeneous domain with no pre-defined defects or heterogeneities. The key difference from the first BVP is that the primary path of the solution induces a heterogeneous stress and strain field. This allows us to test the model's ability to predict spontaneous cohesive instabilities without pre-determined locations of cavity growth. The mechanism driving cavitation here is the varying triaxial stress state as influenced by the boundary conditions. In the primary path of the solution (prior to any instabilities), there is increased shear near the lateral free boundaries that progressively decays as one approaches the center of the specimen. The center of the domain experiences a higher state of tri-axial tension.  Fig.~\ref{fig:strip_contours} illustrates the resulting evolution of the field of nonlocal volume ratio, $\Bar{J}$. The simulation shows that multiple cavities nucleate sequentially, starting from a finite distance from the lateral free boundaries where the specimen still experiences significant shear. As the first pair of cavities forms (taking a characteristic shape reminiscent of the deformed shape of the free boundaries), high levels of shear are taken in the region surrounding the newly formed cavities.  A  boundary-effect from these new "free" boundaries drives a rise of the shear in locations progressively closer to the center, which in turn triggers the nucleation of the next cavities\footnote{The same problem is studied from the perspective of fracture mechanics in \cite{hao2023does}. A more thorough analysis of the range of stress relaxation due to the appearance of new cavity boundaries is needed, especially as it pertains to the strength-surface type criterion that this theory introduces.}. This process repeats, leading to an ordered array of cavities seen propagating toward the center and growing in the sequence of Figs.~\ref{fig:strip_contours}\,(c)-(g). It is postulated that the cavities initiate on the sides as the nucleation criterion considers shear and volumetric contributions in a coupled manner, and a lower threshold is expected at some level of combined volumetric expansion and shear\footnote{This could be calculated by exploring the loss of ellipticity in mixed loading states.}.  This cascade of nucleation events, emerging spontaneously from a homogeneous material, clearly demonstrates the model's ability to capture the complex cavitation patterns. Notably, the undamaged regions in between cavities form a pattern reminiscent of fibrillation \cite{lakrout1999direct, hutchens2016elastic, dorfmann2002shear}, where intact bundles of material are severely deformed and primarily carry the load. In Fig.~\ref{fig:strip_contours}\,(f) extreme loading is leading to the nucleation of new cavities at the base of the formed ``fibrils". This might not be as physically representative, as at these levels of deformation damage, in the form of chain scission is expected to take over.

\begin{figure}[h!]
    \centering
    \includegraphics[width=1\linewidth]{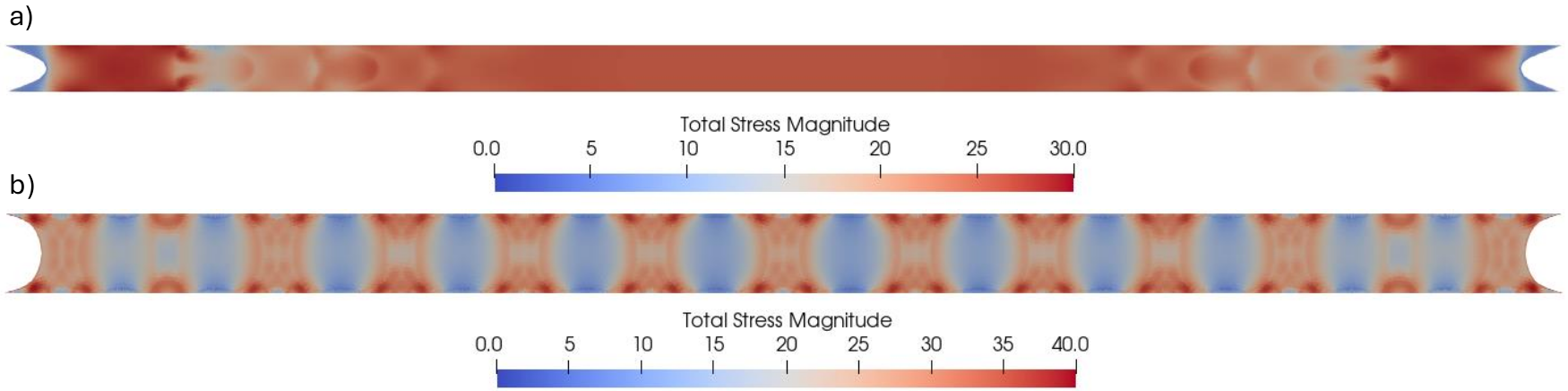}
    \caption{Distribution of the total stress magnitude in the strip. (a) Pre-nucleation state (Step 50) showing high uniform stress. (b) Post-nucleation state (Step 150), illustrating stress relief within the cavities and concentration in the load-bearing ligaments.}
    \label{fig:strip_stress}
\end{figure}

Fig.~\ref{fig:strip_stress} provides the corresponding contours for the magnitude of the total stress at step: (a) 50 and (b) 150, corresponding equivalently to the steps in Fig.~\ref{fig:strip_contours}(b) 50 and (e) 150. As expected, the stress magnitude is lower within the nucleated cavities compared to the surrounding, highly-strained material. However, it is critical to note that the stress does not vanish completely within these cavities, as we have not introduced a damage mechanism. It is expected that the material that has undergone a phase transformation and is in the cavity region will experience a high level of damage, creating dangling chains with no load-carrying capacity that conform to the cavity interface. In the current framework, the nonconvexity of van der Waals free energy leads to a cohesive instability, but as previously discussed, it does not eliminate the volumetric stress entirely, nor does it affect the material's ability to resist shear deformation.  Therefore, to fully examine the evolution of the response and ultimate failure, the extension to incorporate a damage mechanism will be undertaken in Part II.

\begin{figure}[h!]
    \centering
    \includegraphics[width=0.6\linewidth]{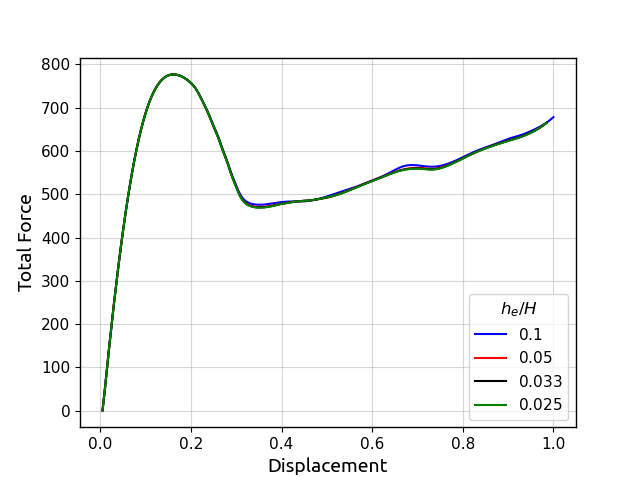}
    \caption{Mesh convergence study for the second BVP using monodisperse network model.}
    \label{fig:strip_mesh_sensitivity}
\end{figure}

A mesh sensitivity study for the pure shear test is presented in Fig.~\ref{fig:strip_mesh_sensitivity}. The total force-displacement curves are plotted for four different element sizes, ranging from $h_e/H = 0.1$ down to $0.025$, while keeping the material length scale $\ell$ fixed. The results indicate convergence, as all four curves are nearly indistinguishable. In this response, after the initial drop due to the first cavity formation, the trend of the curve is upward, indicating a stiffening response due to the fact that the material can still bear loading through fibrillation. In this regard, it should be noted that at displacement around 0.7,  another minor drop in the curve is observed, which is due to the secondary cavitation at the base of the ``fibrils", as previously discussed (seen in Fig. \ref{fig:strip_contours} (f)).

\begin{figure}[h!]
    \centering
    \includegraphics[width=1\linewidth]{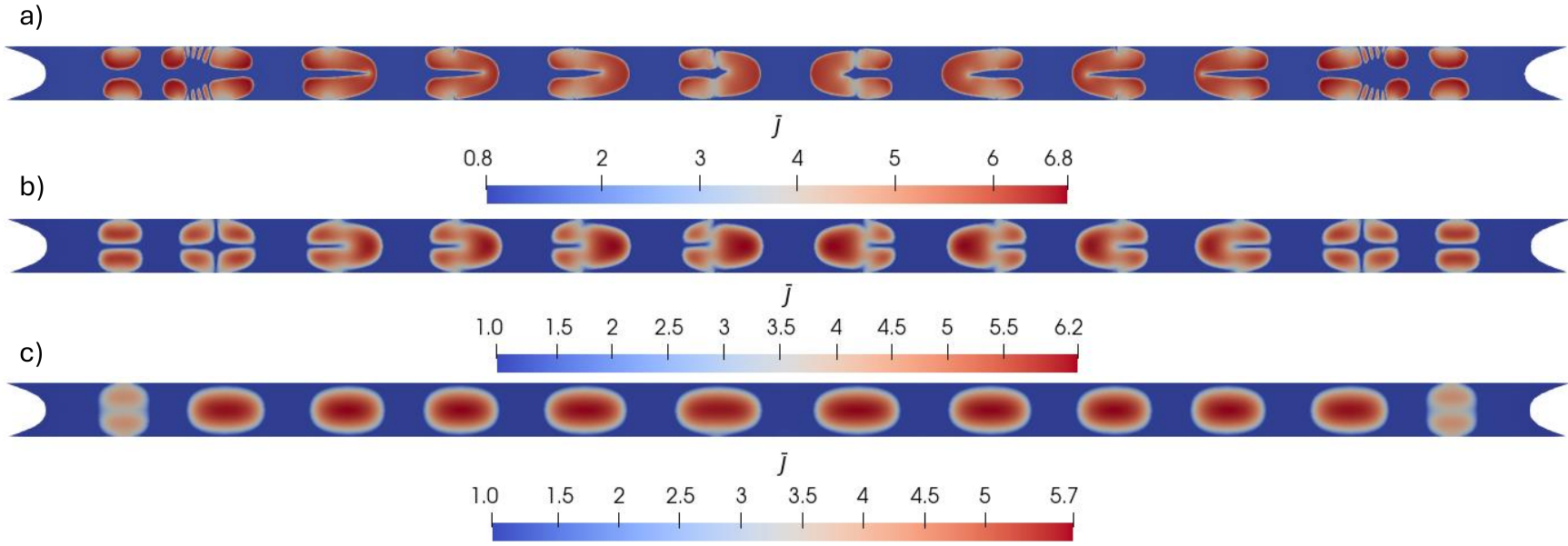}
    \caption{Influence of the nonlocal length scale ($\ell$) on the cavity field morphology in the strip at load step 350. (a) $\ell/H = 0.02$ yields sharp interfaces. (b) $\ell/H = 0.1$. (c) $\ell/H = 0.2$ results in diffuse cavity fields. Note that the location and pattern of cavities remain unaffected by $\ell$.}
    \label{fig:strip_ell}
\end{figure}

\begin{figure}[h!]
    \centering
    \includegraphics[width=0.6\linewidth]{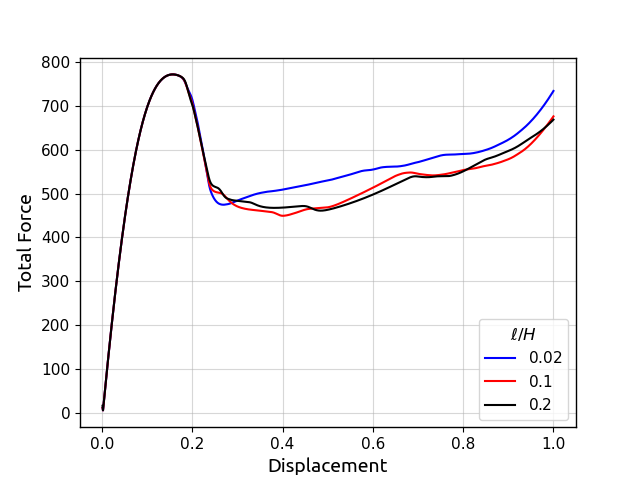}
    \caption{Independence of the macroscopic failure load from the nonlocal length scale $\ell$ in the second BVP. The force-displacement curves are identical up to and including the primary nucleation event (force drop), diverging only during the subsequent propagation phase.}
    \label{fig:strip_ell_graph}
\end{figure}

Finally, we examine the effect of the nonlocal length scale, $\ell$, in the pure shear test. Three values for $\ell/H$ are utilized for testing, namely: 0.02, 0.1, and 0.2. As in the constrained biaxial tension test, this study requires a finer mesh ($h_e/H = 0.02$) to resolve and an increased number of load steps (1000) to properly resolve the gradient field and minimize the influence of viscosity of phase transition.  In Fig.~\ref{fig:strip_ell}, load step 350 (well into the post-bifurcation response) is visualized for all three cases of $\ell/H$. Even though the lower length scales allow for the emergence of finer features in the solution,  the location and pattern of the dominant cavities are surprisingly close for all three length scales. A smaller $\ell$ (Fig.~\ref{fig:strip_ell}\,(a)) produces merging clusters of smaller cavities, while a larger $\ell$ (Fig.~\ref{fig:strip_ell}\,(c)) results in isolated, distinct cavities (except for the cavities near the free boundary). This confirms that $\ell$ controls the size of the features, but the unstable nature of cavitation dominates, with a cascade of cavitation events that control the influence of shear on subsequent cavitation nucleation locations. Moreover, Fig.~\ref{fig:strip_ell_graph} shows that the force-displacement response is entirely independent of $\ell$ up to and including the first unstable drop. The three responses are clearly distinguished during the subsequent expansion and fibrillation phase, even though they all show the same stiffening trend.

\subsection{The effect of aspect ratio in ``pure shear" test}
\begin{figure}[h!]
    \centering
    \includegraphics[width=.7\linewidth]{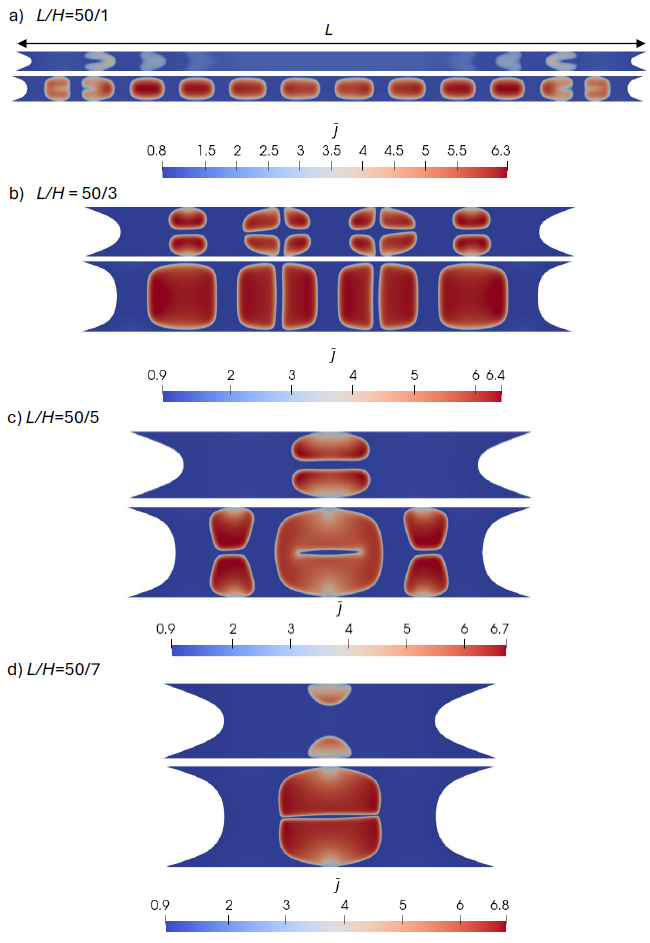}
    \caption{Effect of geometric constraint (aspect ratio $L/H$) on cavitation patterns. (a) The thinnest sample ($L/H=50/1$) shows a dense array of small cavities. (b-c) Intermediate thicknesses. (d) The thickest sample ($L/H=50/7$) shows fewer, massive cavities formed by coalescence.}
    \label{fig:aspect_contours}
\end{figure}

\begin{figure}[h!]
    \centering
    \includegraphics[width=1\linewidth]{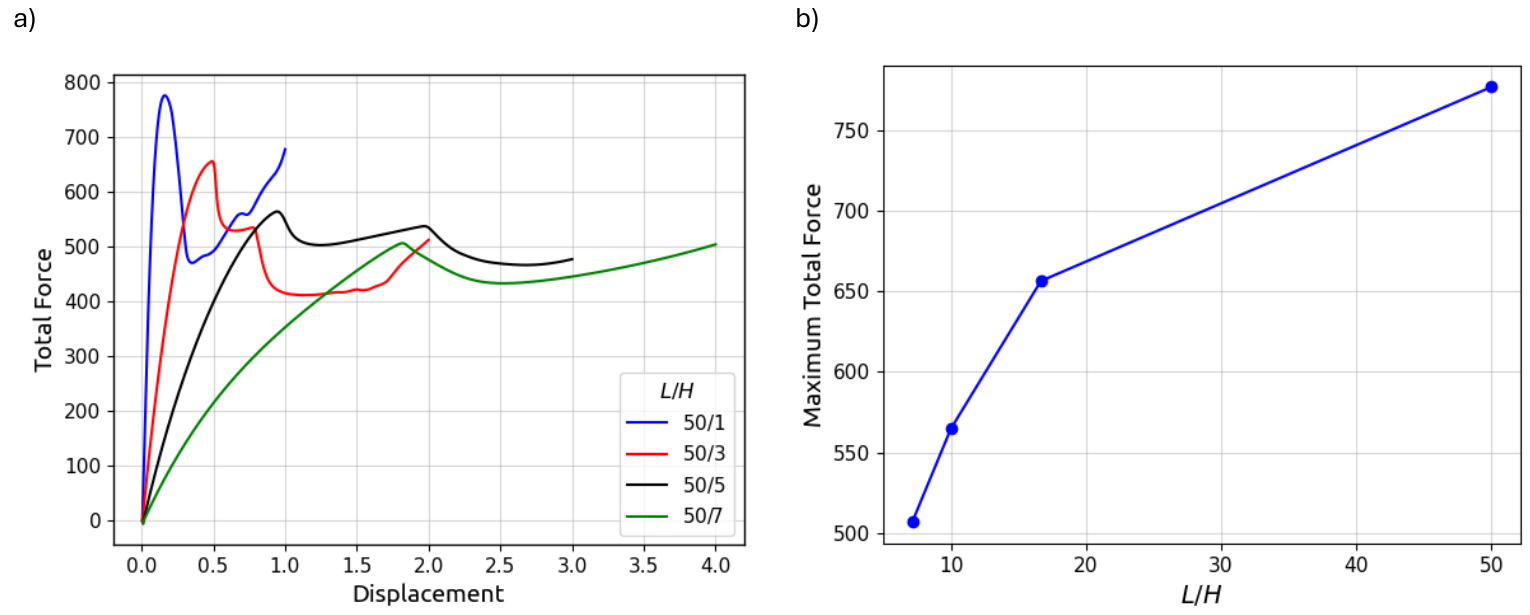}
    \caption{(a) Total force versus displacement curves for varying aspect ratios. (b) Maximum total force as a function of aspect ratio $L/H$.}
    \label{fig:aspect_graph}
\end{figure}

Finally, in the last example of this study, we approach the aspect ratio study for the ``poker chip" test, in a 2D setting, following the pure shear test previously examined.  To this end, assuming the material length scale over the length of the specimen $\ell/L$ is fixed, we change the specimen length-to-height ratio from $50/7$ to $50$, as shown in Fig. \ref{fig:aspect_contours}. The element size and all other parameters are the same for all cases. The only difference with the original pure shear test is the level of maximum displacement, which now depends on the specimen height, as seen in Fig. \ref{fig:aspect_graph} (a). 

First, in Fig. \ref{fig:aspect_contours}, it is shown that the thinner sample forms an array of small cavities, whereas the thicker samples lead to progressively fewer cavities. This observation is in line with experimental observations in \cite{guo2023crack}. In thinner samples ($L/H=50$ and $L/H=50/3$), cavities do not initiate at the center of the specimen. On the contrary, in the thicker samples ($L/H=50/5$ and $L/H=50/7$), the initial cavitation events occur in the vertical centerline of the specimen; specifically in regions closer to the upper and lower boundaries, and then they grow with consequent loading and eventually merge together. This transition for initial cavity formation next to the free boundary for thin samples to the centerline of the specimen for thicker samples is also observed experimentally in \cite{guo2023crack}. The force-displacement curve in Fig. \ref{fig:aspect_graph} (a) indicates that the thinner samples show a stiffer response and earlier cohesive instability and cavity formation. Fig. \ref{fig:aspect_graph} (b) follows the peak values of the applied force versus the specimen aspect ratio. This observation is consistent with the experimental results where increasing the aspect ratio increases the peak load \cite{guo2023crack}. This behavior, as previously discussed, is also expected to arise from the loss of ellipticity in mixed volumetric and shear states that manifest in different BVPs. A more thorough examination of the strength surface resulting from this formulation is needed to provide more clarity regarding the results in these complex stress states.

\section{Conclusion}\label{Section:conclusion}
In this work, we developed a gradient-enhanced continuum framework to model the cohesive instability that can trigger a phase transition from a dense to a rare phase and subsequent cavitation and fibrillation cascade in soft elastomers and adhesives. 
Modeling elastomers as crosslinked van der Waals fluids \cite{lamont2025cohesive}, resistance to volumetric deformations is no longer posed as a mathematical constraint of near incompressibility but directly stems from the micromechanical considerations of the material, leading to a non-(poly)convex Helmholtz free energy for both strain-softening (Neo-Hookean) and strain-stiffening (polymer chain-based) models. This allows 
capturing intermolecular cohesion and excluded volume effects that naturally lead to an unstable response. This creates an alternate view to the problem of cavitation, without requiring a point singularity at the origin of the cavity as introduced in Ball \cite{ball1982discontinuous} or a pre-existing defect \cite{horgan1995cavitation}, but viewing it as a phase transition. Early work of Podio-Giudugli and co-workers has also pointed in this direction \cite{podio1987hyperelastic,lancia1996gleanings}. In this study,  a damage model was not introduced (which will be the focus of part II); as such, the phase transition leads to regions of extreme volumetric deformation, which are interpreted as cavities. More realistically, one expects damage to dominate in these regions and rupture chains to conform to the cavity boundary as dangling chains with no load-carrying capacity.

The proposed theory assumes an excess interfacial energy that can arise during the phase transition, along with viscous effects that are expected from chain disentanglement during cavitation. To allow for a $C^0$ finite element formulation, a nonlocal volume ratio $\Bar{J}$ is introduced, and a thermodynamically consistent derivation of the theory is presented. The model captures the unstable onset of a 2D variant of cavitation in constrained biaxial tension tests, with the instability closely following the analytically predicted threshold. In this setting, a small material heterogeneity is introduced to seed the location of the subsequent phase transition and cavitation. The framework is shown to be mesh insensitive, and the effects of viscous dissipation due to the phase transition, as well as the length scale that controls the interfacial excess energy, are examined. Viscous effects delayed the cohesive instability and cavity formation. Numerical fits to the relationship of the material length scale and referential cavity radius, as well as excess referential interfacial energy, were obtained. A modification in the form of a relaxation function was proposed to enable a constant cavity radius in the reference configuration upon its nucleation.
Utilizing the monodisperse network model enabled linking microstructural parameters such as polymer chain length ($N$) directly to macroscopic response and cavitation resistance. This strain stiffening model showed a very similar response in the constrained biaxial tension test, as the main driver of the instability is the fluid-like van der Waals component of the free energy density. The framework also captured multiple cavity nucleation in a pure shear test (treated as a 2D variant of the poker chip test, but also relevant for thin confined adhesive layers), utilizing a homogeneous specimen without any introduced heterogeneity. In this case,  the model spontaneously captured the sequential nucleation of multiple cavities, driven by the stress state heterogeneity induced by the boundary conditions. This simulation successfully reproduced two experimental observations: (1) the qualitative dependence of failure patterns on aspect ratio, where thin, highly-constrained samples form many small cavities and thick samples form a few large ones (including specifics about their relative location); and (2) the quantitative increase in the peak failure load with decreasing aspect ratio.

While the current formulation focuses on cavity nucleation and growth, it does not yet model chain damage, which will further enable cavity coalescence and ultimately failure, as well as general viscoelastic effects that are extremely relevant for the response of adhesives. On the other hand, it provides a physically grounded foundation for future integration with damage models motivated from polymer chain statistical mechanics (focus of part II) and viscoelastic models relevant for adhesives (focus of part III) to simulate the full cavitation to ultimate failure cascade in soft materials and adhesives.

\section{Acknowledgments}
NB would like to thank Professors Rui Huang and Chung Yuen Hui for early discussions on the topic. 

\appendix

\section{Derivation of Equilibrium State and Bulk Modulus in Plane Strain}
\label{appendix1}

In this work, the material is modeled as a compressible Neo-Hookean solid augmented with a van der Waals fluid contribution. To ensure the simulation begins from a thermodynamically consistent stress-free state for a fixed initial packing fraction $f_0$, we determine the equilibrium volume ratio $J_{eq}$ and the corresponding initial stiffness (Bulk Modulus $K$) under plane strain conditions.

\subsection{Free Energy Density}
The total Helmholtz free energy density $\Psi(J)$ is given by the sum of the entropic (isochoric) and volumetric contributions:
\begin{equation}
    \Psi_{NH-VW} = \Psi_{NH} + \Psi_{VW}
\end{equation}
The compressible Neo-Hookean energy is defined as:
\begin{equation}
    \Psi_{NH} = \frac{\mu}{2}(I_1 - 3 - 2\ln J)
\end{equation}
where $\mu$ is the shear modulus and $I_1 = \text{tr}(\mathbf{C})$ is the first invariant of the right Cauchy-Green deformation tensor. The van der Waals volumetric energy is defined as:
\begin{equation}
    \Psi_{VW} = -\frac{1}{\chi} \left[ \ln(J - f_0) + \epsilon_a \frac{f_0}{J} \right]
\end{equation}
where $\chi$ is the density ratio and $\epsilon_a$ is the normalized attraction energy.

\subsection{Kinematics in Plane Strain}
Under plane strain conditions, the deformation gradient is restricted to the $x_1$-$x_2$ plane. Assuming an equibiaxial pre-stretch $\lambda$, the deformation gradient $\mathbf{F}$ and the volume ratio $J$ are:
\begin{equation}
    \mathbf{F} = \text{diag}(\lambda, \lambda, 1), \quad J = \det(\mathbf{F}) = \lambda^2
\end{equation}
The first invariant $I_1$ relates to $J$ as follows:
\begin{equation}
    I_1 = \lambda^2 + \lambda^2 + 1 = 2J + 1
\end{equation}
Substituting Eq. (A.5) into Eq. (A.2), the elastic energy density in terms of $J$ becomes:
\begin{equation}
    \Psi_{NH}(J) = \frac{\mu}{2}(2J - 2 - 2\ln J) = \mu(J - 1 - \ln J)
\end{equation}

\subsection{Equilibrium Condition}
The hydrostatic pressure $P$ is defined as the conjugate force to the volume ratio $J$. For the material to be stress-free in the initial configuration, the total pressure must vanish:
\begin{equation}
    P = \frac{\partial \Psi_{NH-VW}}{\partial J} = \frac{\partial \Psi_{NH}}{\partial J} + \frac{\partial \Psi_{VW}}{\partial J} = 0
\end{equation}
Differentiating the energy components with respect to $J$:
\begin{align}
    \frac{\partial \Psi_{NH}}{\partial J} &= \mu \left( 1 - \frac{1}{J} \right) \\
    \frac{\partial \Psi_{VW}}{\partial J} &= -\frac{1}{\chi} \left[ \frac{1}{J - f_0} - \frac{\epsilon_a f_0}{J^2} \right]
\end{align}
Combining these, the equilibrium volume $J_{eq}$ is found by solving the following nonlinear equation:
\begin{equation}
    \mu \left( 1 - \frac{1}{J_{eq}} \right) - \frac{1}{\chi} \left[ \frac{1}{J_{eq} - f_0} - \frac{\epsilon_a f_0}{J_{eq}^2} \right] = 0
\end{equation}
For a fixed $f_0$, this equation yields $J_{eq}$, from which the initial pre-stretch is determined as $\lambda_0 = \sqrt{J_{eq}}$.

\subsection{Derivation of the Bulk Modulus}
The Bulk Modulus $K$ represents the material's resistance to volumetric deformation. The stiffness $K$ is the change in pressure ($P$) with respect to the volumetric strain $\epsilon_{vol}$:
\begin{equation}
    K = \frac{dP}{d\epsilon_{vol}}
\end{equation}
Using the chain rule relation $d\epsilon_{vol} = dJ/J$, we can express the derivative as:
\begin{equation}
    \frac{dP}{d\epsilon_{vol}} = \frac{dP}{dJ} \frac{dJ}{d\epsilon_{vol}}
\end{equation}
Substituting $dJ = J \cdot d\epsilon_{vol}$, we obtain the definition used in this study:
\begin{equation}
    K = J \frac{dP}{dJ}
\end{equation}
To evaluate this, we compute the second derivative of the free energy (the derivative of pressure) at the equilibrium state $J_{eq}$. 

For the elastic contribution:
\begin{equation}
    \frac{dP_{NH}}{dJ} = \frac{d}{dJ}\left[ \mu \left( 1 - \frac{1}{J} \right) \right] = \frac{\mu}{J^2}
\end{equation}
For the van der Waals contribution:
\begin{equation}
    \frac{dP_{VW}}{dJ} = \frac{d}{dJ}\left[ -\frac{1}{\chi} \left( \frac{1}{J - f_0} - \frac{\epsilon_a f_0}{J^2} \right) \right]
\end{equation}
\begin{equation}
    \frac{dP_{VW}}{dJ} = \frac{1}{\chi} \left[ \frac{1}{(J - f_0)^2} - \frac{2\epsilon_a f_0}{J^3} \right]
\end{equation}
Finally, the total Bulk Modulus evaluated at $J_{eq}$ is:
\begin{equation}
    K = J_{eq} \left( \frac{\mu}{J_{eq}^2} + \frac{1}{\chi} \left[ \frac{1}{(J_{eq} - f_0)^2} - \frac{2\epsilon_a f_0}{J_{eq}^3} \right] \right)
\end{equation}

\section{Incorporation of a relaxation function in the nonlocal coupling term}\label{appendix2}
Motivated by prior studies \cite{wosatko2021comparison, wosatko2022survey, mousavi2025chain} on damage modeling, we introduce a relaxation function $g(\Bar{J})$ into the nonlocal gradient contribution so that micro–macro nonlocal interactions are progressively attenuated inside the cavity region. The relaxation function can be defined as:
\begin{equation}\label{eq::g-function}
    g(\bar{J}) =
\begin{cases}
1, & \bar{J} \leq \bar{J}_c, \\[6pt]
\exp\!\big(-s \, (\bar{J} - \bar{J}_c)\big), & \bar{J} > \bar{J}_c ,
\end{cases}
\end{equation}
where $\bar{J}_c$ denotes the critical value of volumetric deformation at which cavitation initiates according to the analytical solution, and $s$ governs the rate of exponential decay beyond this point. An example of this function with $\bar{J}_c=1.15$ and $s=0.5$ is illustrated in Fig.~\ref{fig:broadening}.

Physically, this reflects the idea that nonlocal information should not be effectively transmitted from a cavitated zone to the surrounding intact material. The function $g(\Bar{J})$ therefore controls how nonlocal coupling is reduced as the local volumetric deformation increases, governing the decay of nonlocal information transfer from high-volume-change regions toward intact regions.
Incorporating the relaxation function in \eqref{psi_grd-modified} thus alters the Helmholtz free energy, which now takes the form:
\begin{equation}\label{Helmholtz-appendix}
    \psi(\bm{u}, \Bar{J}, \nabla\Bar{J}) = \psi(\bm{u}) -\frac{nk_{b}\theta}{\chi}\left[\ln(\Bar{J}-f_{0})+\frac{\epsilon_{a}}{k_{b}T}\frac{f_{0}}{\Bar{J}}\right] + c \left[J - \Bar{J}\right]^2 + \frac{\ell^2}{2} g(\Bar{J}) \nabla\Bar{J} \cdot \nabla\Bar{J}
\end{equation}
which also changes the constitutive laws as follows:

\begin{equation}\label{constitutives-appendix}
\begin{split}
    \boldsymbol{\xi}_{\Bar{J}} = \ell^2g(\Bar{J})\nabla\Bar{J}\quad \text{in} \quad \Omega_0, \\
    f_{\Bar{J}} =  -\frac{nk_{b}\theta}{\chi}\left[\frac{1}{\Bar{J}-f_0}-\frac{\epsilon_{a}}{k_{b}T}\frac{f_{0}}{\Bar{J}^2}\right] -\left[J - \Bar{J}\right] + \frac{\ell^2}{2} \frac{\partial g(\Bar{J})}{\partial \Bar{J}} \nabla\Bar{J} \cdot \nabla\Bar{J} \quad \text{in} \quad \Omega_0.
\end{split}
\end{equation}


\end{document}